\documentclass{jpp}
\usepackage{graphicx}
\usepackage{url}

\usepackage[utf8]{inputenc}
\usepackage[T1]{fontenc}
\usepackage{amsmath}

\shorttitle{Heat conduction in magnetised, weakly collisional plasma}
\shortauthor{T. A. Vincent and others}

\title{Design of experiments characterising heat conduction in magnetised, weakly collisional plasma}

\author{T. A. Vincent\aff{1}
  \corresp{\email{thomas.vincent@physics.ox.ac.uk}},
  P. Ariyathilaka\aff{2}, L. Creaser\aff{3}, C. Danson\aff{3}, D. Lamb\aff{4}, J. Meinecke\aff{5}, C. A. J.  Palmer\aff{6}, S. Pitt\aff{3}, H. Poole\aff{8}, C. Spindloe\aff{2}, P. Thomas\aff{3}, E. Tubman\aff{7}, L. Wilson\aff{3}, W. J. Garbett\aff{3},  G. Gregori\aff{1}, P. Tzeferacos\aff{8}, T. Hodge\aff{3}, A. F. A. Bott\aff{1}}

\affiliation{\aff{1}Department of Physics, University of Oxford, Parks Rd, Oxford OX1 3PU, UK
\aff{2}Central Laser Facility, Rutherford Appleton Laboratory, Didcot OX11 0QX, UK
\aff{3}AWE, Aldermaston, Reading, West Berkshire, RG7 4PR, UK
\aff{4}Department of Astronomy and Astrophysics, University of Chicago, Chicago, IL 60637
\aff{5}Department of Physics, Gettysburg College, Gettysburg, PA 17235
\aff{6}School of Mathematics and Physics, Queen’s University Belfast, Belfast BT7 1NN, UK
\aff{7}Department of Nuclear Engineering, University of California Berkeley, California, CA 94720
\aff{8}Department of Physics and Astronomy, University of Rochester, Rochester, NY 14627 }

\begin{document}

\maketitle

\begin{abstract}
Heat conduction in weakly collisional, magnetised plasma is challenging to model accurately due to multifaceted physics governing heat-carrying electrons, including microinstabilities that scatter electrons and modify heat transport. Capturing these effects requires multidimensional kinetic theory simulations, which are computationally expensive. Experimental constraints overcome this issue, resulting in improved understanding of thermal transport in systems such as the intra-cluster medium of galaxy clusters, and the hot-spot in inertial confinement fusion. In this paper, we present a new experimental platform that produces a weakly collisional high-$\beta$ plasma expected to be susceptible to the whistler heat-flux instability. This platform, to be fielded on the Orion laser, enables characterisation of whistler-regulated thermal conductivity. The platform design is assessed using radiation-magnetohydrodynamics simulations with the code FLASH. Simulations using three thermal conduction models predict conductivity suppression by over an order of magnitude relative to the Spitzer value at whistler saturation, demonstrating the efficacy of the platform.
\end{abstract}

\section{Introduction}
\label{sec:intro}
Thermal transport is an important process in a range of plasma environments, ranging from high energy density physics (HEDP) and inertial confinement fusion (ICF) plasmas to the intracluster medium (ICM) of galaxy clusters. In the context of ICF, thermal conduction plays a pivotal role in the formation and evolution of hot-spot, with heat conduction from the hot-spot to the surrounding cooler dense plasma one of the primary loss mechanisms during burning fuel-capsule implosions~\citep{abu2022lawson}. Accurate modelling of heat transport is therefore essential for realising successful ICF experiments; yet classical models exhibit systematic discrepancies when compared with observations. Such discrepancies point to the role of kinetic processes~\citep{rinderknecht2018kinetic}, while magnetic fields generated prior to the stagnation phase of ICF implosions may also significantly influence heat transport~\citep{walsh2017self, walsh2021biermann}. Recent studies have explored how imposed magnetic fields during ICF implosions could positively alter the properties of the fusing plasma~\citep{o2025burn,Walsh_2025}. For instance, introducing a magnetic field across the fuel-capsule prior to implosion has been shown to increase ion temperatures by up to $40\%$ and neutron yields by more than threefold, relative to unmagnetised shots of similar drive energy \citep{chang2011fusion, moody2022increased}. These results suggest that external magnetic fields can significantly enhance the performance of ICF experiments.

Thermal conduction is also thought to play a key role in one of the longest-standing puzzles in the physics of the ICM: the cooling-flow problem~\citep{fabian1994cooling}. Observed luminosity and density profiles from galaxy cluster cores and the surrouding ICM suggest that radiative cooling times should be much shorter than the Hubble time: a discrepancy that has been a central issue in the field for several decades \citep{lea1973thermal, fabian1977subsonic, cowie1977radiative}. Such radiative losses would, in principle, drive mass accretion onto the cooling cores to maintain pressure balance with the gravitational potential of the surrounding gas. This could initiate a feedback loop in which radiative core cooling is followed by the inflow of cooling material onto the cluster's core. The result would be cooling instabilities which in turn would yield mass deposition and stellar formation rates that are much greater than what is observed. Despite evidence of cooling flows, key dynamical properties of clusters with cooling cores -- for example, the mass deposition rate, central cooling time, and central entropy -- have remained approximately constant over several Gyr \citep{mcdonald2013growth, mcdonald2018revisiting}. The lack of evolution indicates a need for a thermal balance to keep the cluster cores and ICM in equilibrium. Active galactic nuclei (AGN) feedback from the cluster galaxies is thought to be the dominant source of this heating \citep{fabian2012observational, ruppin2023redshift}, with the role of thermal conduction remaining a challenge in understanding the thermodynamics of the ICM. It has been predicted that thermal conduction needs to be suppressed to prevent cool-cores from isothermalising without the need for fine tuning \citep{conroy2008thermal}, with the observation of persistent cold fronts similarly requiring this \citep{zuhone2016cold}. The presence of tangled magnetic field lines within the ICM suggests that various mechanisms could underlie this suppression \citep{kempf2025non}, with significant implications for the efficacy of thermal balancing attributed to thermal conduction into the cluster core~\citep{kunz2011thermally}. 

In collisional plasmas where the Coulomb mean free paths of electrons and ions, $\lambda_{e}$ and $\lambda_{i}$, are much smaller than the characteristic temperature length scale $L_{\rm T}$, it is possible to derive analytical expressions for the thermal conductivity using kinetic theory: so-called `classical models'. In unmagnetised plasmas where the electron Larmor frequency $\Omega_{e}$ and the electron-ion collision time $\tau_{ei}$ satisfy $\Omega_{e} \tau_{ei} \ll 1$, such calculations yield the Spitzer model, in which heat flows down temperature gradients in the plasma, carried predominantly by electrons~\citep{spitzer1953transport, cohen1950electrical}. The Spitzer heat flux $q_{S}$ is directly proportional to $\tau_{ei}$, and has a strong dependence on the electron temperature: ${q_{S} \sim m_e n_e v_{\mathrm{th}e}^4 \tau_{ei}/L_{\rm T}\propto T_{e}^{7/2}}$, where $m_e$ is the electron mass, $n_e$ the electron number density, $v_{\mathrm{th}e}$ the thermal electron velocity, and $T_{e}$ the electron temperature. In a collisional plasma with a temperature gradient that is strongly magnetised ($\Omega_{e}\tau_{ei} \gg 1$), it can be shown that, compared with its parallel component $q_{||}$, the heat flux in the direction perpendicular to the magnetic field $q_{\perp}$ will be strongly suppressed: $q_{\perp}/q_{||} \sim (\Omega_{e}\tau_{ei})^{-2} \ll 1$ \citep{braginskii1965transport}. The parallel heat flux is still given by the Spitzer model, with $q_{||} \propto T_{e}^{7/2}$. 

However, the plasmas present in both ICF hot-spot and the ICM are only weakly collisional, with important ramifications for modelling thermal transport. The Coulomb mean free paths $\lambda_e$ and $\lambda_i$ are only a few orders of magnitude smaller than $L_{\rm T}$, and their thermal pressure $p$ is much larger than their magnetic pressure $p_{B}$, resulting in a large plasma $\beta \equiv p/p_{B}$. Plasmas with these two characteristics, which also include other notable astrophysical and HED settings such as the warm intergalactic medium \citep{nicastro2008missing} and laser-ablated plasma \citep{tzeferacos2018laboratory, bott2021time, meinecke2022strong}, are expected to be susceptible to anisotropy-driven kinetic-scale instabilities when $\lambda_e/L_{\rm T} \gtrsim 1/\beta$~\citep{bott2024kinetic}. These instabilities mean that classical models for heat conduction in magnetised plasma, which assume that heat transport is mediated only by Coulomb collisions, are inaccurate when the plasma is weakly collisional with a sufficiently large plasma $\beta$. 

During the last few decades, various new models of thermal conductivity that incorporate the effect of kinetic instabilities have been put forward. \citet{Levinson_1992} proposed that the whistler heat-flux instability, in which the temperature-gradient-induced heat-flux acts as a source of free energy for whistler waves to grow, could affect heat transport once their amplitude is large enough to scatter the pitch-angle of heat-carrying electrons. Various analytical~\citep{Pistinner_1998,Drake_2021} and numerical studies~\citep{RobergClark_2016,komarov2018self, roberg2018suppression, yerger2025collisionless} have carried out further research on the instability,
confirming that anomalous scattering causes a reduction in the parallel heat flux relative to the Spitzer value. For example, \citet{komarov2018self} report that
\begin{equation}
    \frac{q_{||\text{eff}}}{q_{S}} = \left[ 1 + \frac{\beta_{e}}{3\left( L_{\rm T}/\lambda_{e} + 4 \right)} \right]^{-1} .
    \label{eq:komarovHeatFlux}
\end{equation}

This expression indicates that if $\lambda_{e}/L_{\rm T} \gg 1/\beta_e$, the parallel heat flux will be strongly suppressed, and become independent of the magnitude of the temperature gradient. Another model, appropriate for any kinetic instability that generates magnetic fluctuations with an amplitude comparable to the background magnetic field strength, was proposed by~\citet{ryutov1999similarity}: $q_{||\text{eff}} \sim q_{\perp\text{eff}} \sim \ell_{\delta B} m_e n_e v_{\mathrm{th}e}^3/2 L_{\mathrm{T}} \sim  q_{S}\ell_{\delta B} /\lambda_{e}$, where $\ell_{\delta B}$ is the correlation scale of the magnetic perturbations. In this model, conduction is quasi-isotropic, even though the Larmor radius $\rho_e \ll \lambda_e$, but is still proportional to the temperature gradient. Although in specific context these models have been validated \citep{yerger2025collisionless}, their applicability more generally in weakly collisional plasmas remains uncertain.   

Advancements made in kilo-joule to mega-joule, and peta-watt laser facilities \citep{danson2019petawatt} such as the National Ignition Facility (NIF), and the OMEGA laser at the Laboratory for Laser Energetics (LLE) have made laboratory investigations of heat transport in weakly collisional, magnetised plasma possible. Although experiments investigating the effect of nonlocality~\citep{henchen2018observation} and magnetic fields~\citep{Froula_2007} on heat transport in laser plasmas are well established, recent experiments reported by~\citet{meinecke2022strong} represent a significant step forward in this regard, providing evidence for suppressed heat conduction compared with classical models in a turbulent, magnetised, weakly collisional plasma. In these experiments, which were carried out on the NIF, temperature fluctuations of characteristic scale  $L_{T} \gtrsim 5 \lambda_{e}$ were observed to persist for hundreds of classical thermal diffusion times, consistent with a suppressed conduction compared to the Spitzer model by around two orders of magnitude.  
This suppression was linked to electron magnetisation by 
comparing the NIF experimental results with the conduction characteristics of turbulent plasmas created in experiments at the Omega Laser Facility~\citep{tzeferacos2018laboratory,bott2021time} which, besides smaller values of $\Omega_e \tau_{ei}$ comparatively, were physically similar. Suppressed conduction was only observed when $\Omega_e \tau_{ei} \gg 1$, the NIF produced plasma; experiments in which $\Omega_e \tau_{ei} \lesssim 1$ yielded smooth temperature profiles despite exhibiting comparable turbulent velocity statistics. However, beyond this correlation, the suppression mechanism was not well constrained, with various different possibilities plausibly operating: non-local effects, the trapping of heat-carrying electrons by magnetic mirrors \citep{komarov2016thermal}, and predominantely orthogonal arrangements of temperature gradients and magnetic fields~\citep{komarov2014suppression} as well as whistler heat-flux instabilities. Testing specific conduction models in such a complex plasma environment -- in particular, one with a highly stochastic magnetic field -- is a significant challenge.     

To overcome this challenge, we have designed a new experimental platform to be fielded on the Orion laser facility \citep{hopps2015comprehensive}, that produces a magnetised, high-$\beta$, weakly collisional plasma with temperature gradients directed along embedded magnetic field lines with a simple geometry. The goal of this platform is to characterise thermal conduction in a plasma susceptible to whistler heat-flux instabilities for comparison against a range of different heat conduction models. This paper contains a description of the new experimental platform, as well as a series of complementary simulations using the radiation-magnetohydrodynamics (MHD) code FLASH \citep{fryxell2000flash}, and also synthetic diagnostic data. The simulations were run with various conduction models supported by FLASH to investigate how the evolution of the plasma is affected by thermal conductivity. Our key finding is that the evolution of the temperature profile is dependent primarily on the heat conduction model, suggesting the temperature measurements will provide significant constraints on the thermal conductivity. 

The remainder of this article is structured as follows. Section \ref{sec:design_overview} describes the design overview of the platform, with descriptions of the target and previous iterations of the design alongside FLASH simulation outputs of the resulting plasma. Section \ref{sec:diagnsotics} discusses the three primary diagnostics to be fielded throughout the campaign, with analysis of synthetic data produced for each diagnostic. Finally, section \ref{sec:discussion} discusses the evolution of a set of notable plasma parameters relevant to the problem, and how the thermodynamics compare between the three conduction models.

\section{Design overview}
\label{sec:design_overview}
\begin{figure}
    \centering
    \includegraphics[width=1\linewidth]{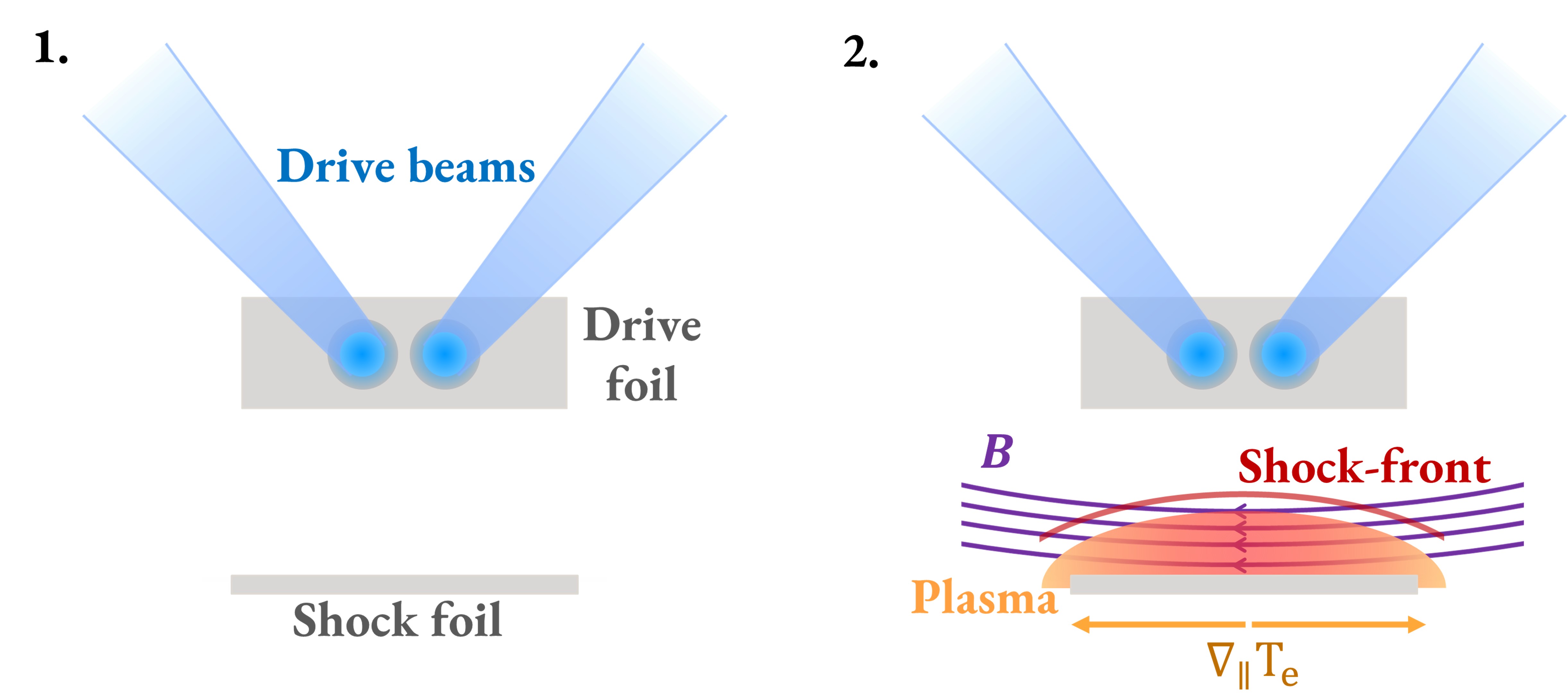}
    \caption{Simplified diagram of plasma generation in two phases: the driving of the target foil producing a front-side blow-off plasma. Then, the generation of a shocked plasma from the front-side blow off, threaded with self-generated magnetic fields parallel to the temperature gradients generated during the drive foil ablation.}
    \label{fig:cartoon_experiment}
\end{figure}

To produce a magnetised, weakly collisional, high-$\beta$  plasma with temperature gradients directed along magnetic field lines such that the whistler heat-flux instability can grow, we propose the experimental design illustrated in figure. \ref{fig:cartoon_experiment}. A front-side blow-off plasma is produced by ablating a foil with multiple drive lasers, expanding from the target in a hemispherical plume with radius parallel to the foil normal. The supersonic blow-off plasma then collides with a secondary `shock' foil with its plane face normal to that of the drive foil. The resulting shock wave in the plasma jet will convert the majority of its kinetic energy into thermal energy, resulting in a subsonic planar plasma \citep{nagayama2000theory, colvin2013extreme}.  Pressure gradients in the central blow-off plume, which are tuned by the laser spot separation, generate temperature gradients from the central dense planar region to the less dense wings toward the edges of the shock foil. The plasma's magnetic field is generated by the Biermann battery mechanism operating on misaligned density and temperature gradients present in the front-side blow-off plasma~\citep{biermann1950ursprung, stamper1971spontaneous, stamper1975faraday}. This plasma has a large magnetic Reynolds number, so the magnetic field is advected with the blow-off plasma as it expands~\citep{alfven1943existence}. Although this expansion also dilutes the magnetic field strength, as the field passes through the shock wave, its component perpendicular to the shock normal is re-amplified, resulting in a magnetic field in the planar plasma whose orientation is primarily in the plane of the shock foil.

The temperature, density, magnetic field strength, and temperature gradient scale $L_{\rm T}$ in the planar plasma can be tuned to access the regime of interest by varying the intensity of drive lasers, their spot profile, and the separation of the drive and shock foils. For the whistler heat-flux instability to operate at all, we require that $L_{\rm T}/\lambda_{e}\lesssim\beta_{e}$ within the planar plasma, and the Hall parameter $\mathrm{Ha}_{e} \equiv \lambda_{e}/\rho_{e}$ to satisfy $\mathrm{Ha}_{e} \gtrsim 1$. We also require that both the characteristic e-folding timescale $\tau_{\rm lin} \sim L_{\rm T}/\lambda_{e}\Omega_e$ of the whistler heat-flux instability and the timescale $\tau_{\text{c}}\sim L_{T}/\beta_{e} v_{\mathrm{th}e}$ over which the amplified whistlers scatter electrons to be much smaller than the dynamical timescale $\tau_{\text{dyn}}$ over which the temperature gradient evolves~\citep{komarov2018self, bott2024kinetic}. To ascertain under which platform design parameters the relevant conditions were obtained, we carried out several simulation campaigns using the FLASH plasma-fluid simulation code. FLASH \citep{fryxell2000flash} is a publicly available Eulerian, finite-volume radiation-MHD code with extensive HEDP capabilities \citep{tzeferacos2015flash}, making use of equation of state (EoS) and opacity tables from PrOpacEoS \citep{macfarlane2006}. This makes it a reliable tool for the simulation of laser-plasma experiments, validated against a range of high-power laser-plasma experiments \citep{Fatenejad2013172, tzeferacos2015flash, Tzeferacos2017, tzeferacos2018laboratory, albertazzi2018experimental, rigon2019rayleigh, bott2021time, bott2021inefficient, bott_2022, moczulski2024numerical}.

\subsection{Target design} \label{sec:target_design}

The initial phase of the target's design process involved finding appropriate choices for the characteristics of the drive and shock foils, as well as a long-pulse (LP) beam geometry appropriate to be fielded on the Orion Laser Facility~\citep{hopps2015comprehensive}. Each LP beam on Orion delivers up to $400$J at a wavelength of $351$ nm. A previous experiment carried out on the facility focused each LP beam onto separate $\sim$$250 \; \mu$m diameter spots on chlorine-doped millimetre-scale CH targets using a $1$ ns square pulse and obtained weakly collisional front-side blow-off plasmas with values of $\beta \sim 1$-$100$ and $\mathrm{Ha}_e \sim 10$~\citep{tubman2021observations}. Such values are appropriate for our proposed study, suggesting that irradiating millimetre-scale drive and shock chlorine-doped CH foils using Orion's LP beams in a similar manner would be a productive basis for the platform. 

\begin{figure}
    \centering
    \includegraphics[width=1\linewidth]{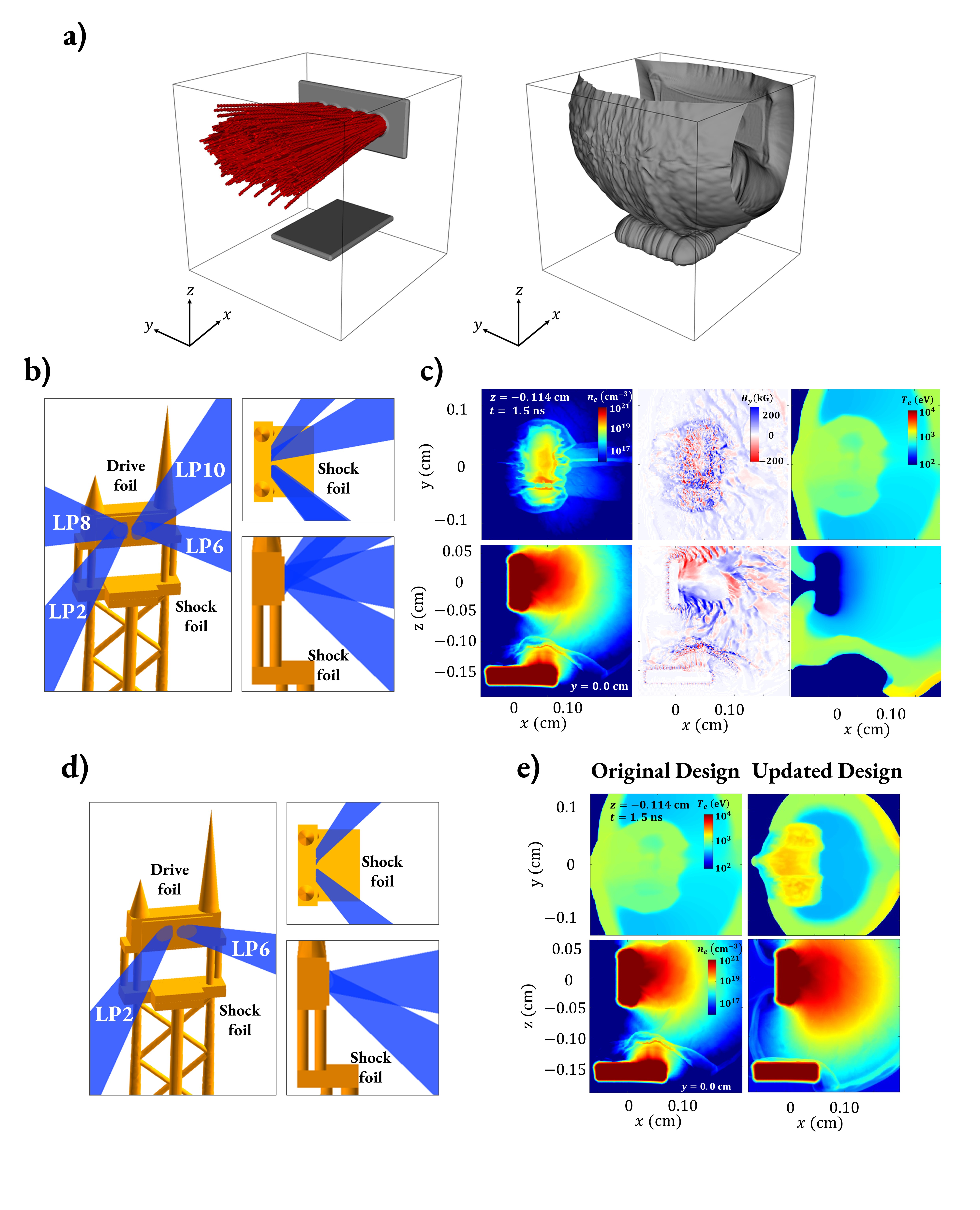}
    \caption{Evolution of the target design. (a) Original five beam proof of concept visualisation with all drive beams normal to target. The left shows all beams incident on the drive foil, and the resultant ablation plume on the right, with end of the shock foil ablating prior to contact with the jet plasma. (b) Updated four beam design visualisation using the Orion chamber beam alignment geometry, and shorter shock foil. The right top panel shows the target top-down, and the bottom right panel showing the target side-on matching the LoS seen in the top and bottom row, respectively, of (c). (c) 2D slices taken from the FLASH simulation data for $T_{e}$ (left), $B_{y}$ (centre), and $n_{e}$ (right), for $z = -0.114$ cm (top row), and $y = 0.00$ cm (bottom row) at $t = 1.5$ ns post the drive foils ablation. (d) Same as (b), but with the final two beam configuration (e) Comparison of 2D slices from FLASH simulation data for $T_{e}$ (top row), and $n_{e}$ (bottom row), as in (c) for the two beam and four beam design, indicating the more laminar nature of the temperature and density profiles for the two beam design.}
    \label{fig:targetHistory}
\end{figure}

To refine the design, a series of three-dimensional (3D) FLASH simulations were performed, investigating different configurations. In each of these simulations, the appropriate material properties of chlorine-doped CH targets was used, as well as the LP beam parameters described in the previous paragraph (see figure \ref{fig:targetHistory}). 
During the campaigns, the simulation domain was a uniform, $256^{3}$ Cartesian grid of resolution $\Delta x = \Delta y = \Delta z = 1\times10^{-3}$ cm. Here, the $x$-direction is that normal to the drive foil, the $z$ direction normal to the shock foil, and the $y$ direction the third orthogonal direction. The origin of the domain was centred such that $-0.2021$ cm $\leq x \leq 0.0721$ cm, $-0.1276$ cm $\leq y \leq 0.1276$ cm, and $-0.2021$ cm $\leq z \leq 0.0721$ cm. Outflow boundary conditions were utilised in all simulations. The first design used five drive beams directed normal to the plane of the drive foil (figure \ref{fig:targetHistory}a), and a shock foil that extended 2 mm in the direction anti-parallel to the drive beam propagation axis. This produced a weakly collisional, high-$\beta$ plasma with parameters generally compatible with operation of the whistler heat-flux instability. However, the extended length of the shock foil caused issues: partial reflection and scattering of the drive beams by the front-side blow-off plasma ablated its end (figure \ref{fig:targetHistory}a). This ablation did not  affect the planarity of the shocked plasma directly below the drive foil. However, magnetic fields oriented opposite to those in the bulk of the planar plasma were produced in the ablated shock foil, significantly distorting path-integrated measurements of magnetic fields. Because the extended length yielded no apparent benefits, a new design with a shorter shock foil was an obvious improvement. 

Using the shortened platform, the beam profiles and angles of incidence were then updated to more accurately represent the capabilities of the Orion facility \citep{hopps2015comprehensive}. The number of beams was reduced to four because of geometric feasibility, making use of LP beams 2, 6, 8, and 10 (see figure \ref{fig:targetHistory}b). This update highlighted the importance of accurate modelling of the angle of incidence of the drive beams with respect to the drive foil. The incidence of LP-8 and 10 onto the drive foil from above the plane of its normal led to worsened scattering of laser light onto the shock foil compared with previous five-beam design, a feature that even the shortened platform could not avoid. Although the resulting plasma was again weakly collisional, and high-$\beta$, this scattered light generated non-planar features in the temperature and density profiles of the shocked plasma, as well as pronounced stochasticity in its magnetic fields (figure \ref{fig:targetHistory}c). Both features prohibited use of LP-8 and LP-10 in this platform. 

To avoid this issue, a new two-beam design using only LP-2 and 6 was considered (figure \ref{fig:targetHistory}d). This design avoided any ablation of the drive foil by scattered laser light, resulting in a genuinely planar, weakly collisional, high-$\beta$ plasma suitable for our experiment of interest (figure \ref{fig:targetHistory}e). In this new design, the separation $L_{\rm spots}$ between the spots was then tuned to control the horizontal size of planar plasma, while avoiding any magnetic reconnection effects in the drive foil (see section \ref{sec:sim_plasma_param}). The distance $L_{\rm shock}$ of the shock foil from the drive foil was chosen via a simple optimisation: too close, and the curvature of the blow-off plasma  causes departures from planarity; too far, the plasma density becomes too low to be easily detectable. 

\begin{figure}
    \centering
    \includegraphics[width=1\linewidth]{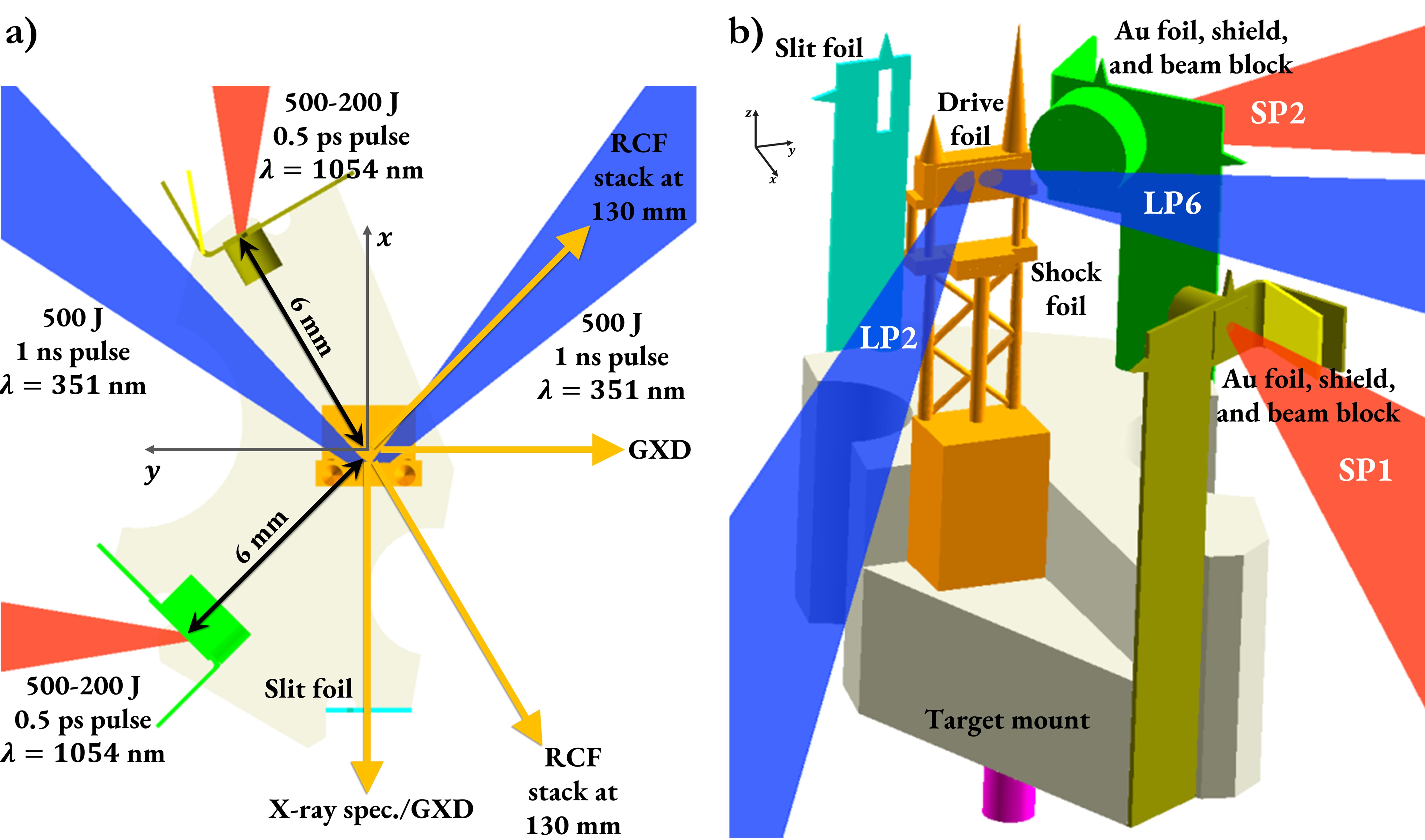}
    \caption{(a) VisRad visualisation of the experimental configuration within the Orion chamber. The drive and shock foils are placed in the centre of the chamber with the plane face of the shock foil facing out of the page, and the plane face of the drive foil facing northward. The LP drive beams are shown in blue targeting the drive foil, with the SP beams shown in red targeting the Au foils for proton production. The diagnostic detectors are off-screen with their radial position, relative to target chamber centre (TCC), and lines-of-sight (LoS) indicated by the orange arrows. The slit foil at the bottom of the image is to screen the X-rays for the spectrometer, which will be swapped out mid-way through the campaign to be replaced with a spatially resolved GXD. (b) Target configuration on Al mounting block. The orange chair will be 3D printed with with the drive and shock foils mounted on it. The cross-beam structure holding the chair will increase stability and ensure that the drive and shock foils' plane faces will remain orthogonal post ablation.}
    \label{fig:targetSchematic}
\end{figure}

A detailed visualisation of the target to be fielded with the two-beam design is shown in figure \ref{fig:targetSchematic}.
The drive and shock foils consist of $1.6$ mm $\times$ $0.75$ mm $\times$ $0.2$ mm and $1.2$ mm $\times$ $0.75$ mm $\times$ $0.2$ mm  slabs of parylene C ($50\%$C, $43.75\%$H, $6.25\%$Cl) and plastic ($50\%$C, $50\%$H), respectively, separated by a $0.8$ mm gap between the bottom of the drive foil and the top face of the shock foil. To maximise target stability, the foils are slotted into a 3D printed `chair', in turn mounted onto a machined aluminium (Al) block. 

In addition to the main target, various additional components are mounted onto the Al block, to facilitate diagnostics. The mounting block supports an adjustable $10.6$ mm $\times$ $2.0$ mm $\times$ $0.2$ mm Al sheet, containing a $0.8$ mm $\times$ $0.3$ mm vertical slit, to set the resolution of the x-ray spectrometer (see section \ref{sec:spectroscopy}). This sheet is positioned with its plane face normal to the rear side of the target chair, $5.85$ mm away. It is set such that only the area between the drive and shock foils is visible to the spectrometer. Additionally, two circular gold (Au) foils are fastened to the mounting block, from which proton beams for imaging will be generated using Orion's short pulse (SP) beams. The foils are 0.05 mm thick, have $1.1$ mm and $2.4$ mm diameters, and are 
$6$ mm away from the centre of the chair, at angles of $30^{\circ}$ and $135^{\circ}$ counter clockwise relative to the normal of the drive foil's face, respectively. They are centred at $0.15$ mm above the shock foil. Protective cylinders, $1$ mm and $2$ mm long with an inner diameter of $1$ mm, are used to shield the Au foils to prevent pre-ablation to the foil surface for the $30^{\circ}$ and $135^{\circ}$ foils. Beam blockers are mounted on the frame containing the Au foils, placed at an angle of $68^{\circ}$ normal to the foil plane, to prevent reflection of the SP beam into the parabola of the adjacent SP beam. The expected on-target intensity of the drive beams is $6\times10^{14}$ W cm$^{-2}$ horizontally separated by $0.5~\text{mm}$. The SP beams each deliver $5\times 10^{20}$ W cm$^{-2}$ to their respective target, containing up to $500$ J over a $5\times10^{-4}$ ns pulse at a wavelength of $1054$ nm. The LPs will have super-Gaussian beam profiles, shaped by phase plates to create circular on-target profiles with $230~\mathit{\mu}\text{m}$ focal spot size, while the SPs have an approximate focal spot size of $7.7~\mathit{\mu}\text{m}$.

\subsection{Simulated plasma parameters} \label{sec:sim_plasma_param}

One key benefit of designing a new experimental platform using a plasma simulation code such as FLASH is the possibility of detailed characterisations of the plasma conditions realised, and their sensitivity to different models of conduction. In this section, we explore the evolution of temperature, density, and magnetic fields in three different simulations of the final two-beam design that employed different conduction models: flux-limited Spitzer, Ryutov, and no conduction. When the specifics of heat conduction are uncertain, simulations with a range of models, including the most discrepant ones, are essential for exploring the range of possible experimental outcomes. A simulation using flux-limited Spitzer is one useful benchmark: as theoretical expectations are that conduction should be suppressed compared to this model, simulations using it can be considered to represent the "maximal" conduction scenario. By contrast, the ``conduction-off" simulation represents the opposite scenario. Finally, the Ryutov model is a particularly simple `modified' conduction model to use numerically -- unlike the Komarov model, its heat flux is not anisotropic with respect to the macroscopic magnetic field -- making it an attractive third option. In this set of simulations, only the drive and shock foils were included, for computational efficiency. The simulation parameters are the same as those described in section \ref{sec:target_design}. It should be noted that kinetic effects are not included in FLASH, which could lead to some systematic uncertainty that should be considered when comparing with experimental data. The ion-Coulomb mean-free path $\lambda_{i}$ is larger than the shock-front for these simulations, indicating that collisionless effects may be present around the shock. This could imply a slight underprediction in the densities given by FLASH when the region first forms.

\begin{figure}
    \centering
    \includegraphics[width=1\linewidth]{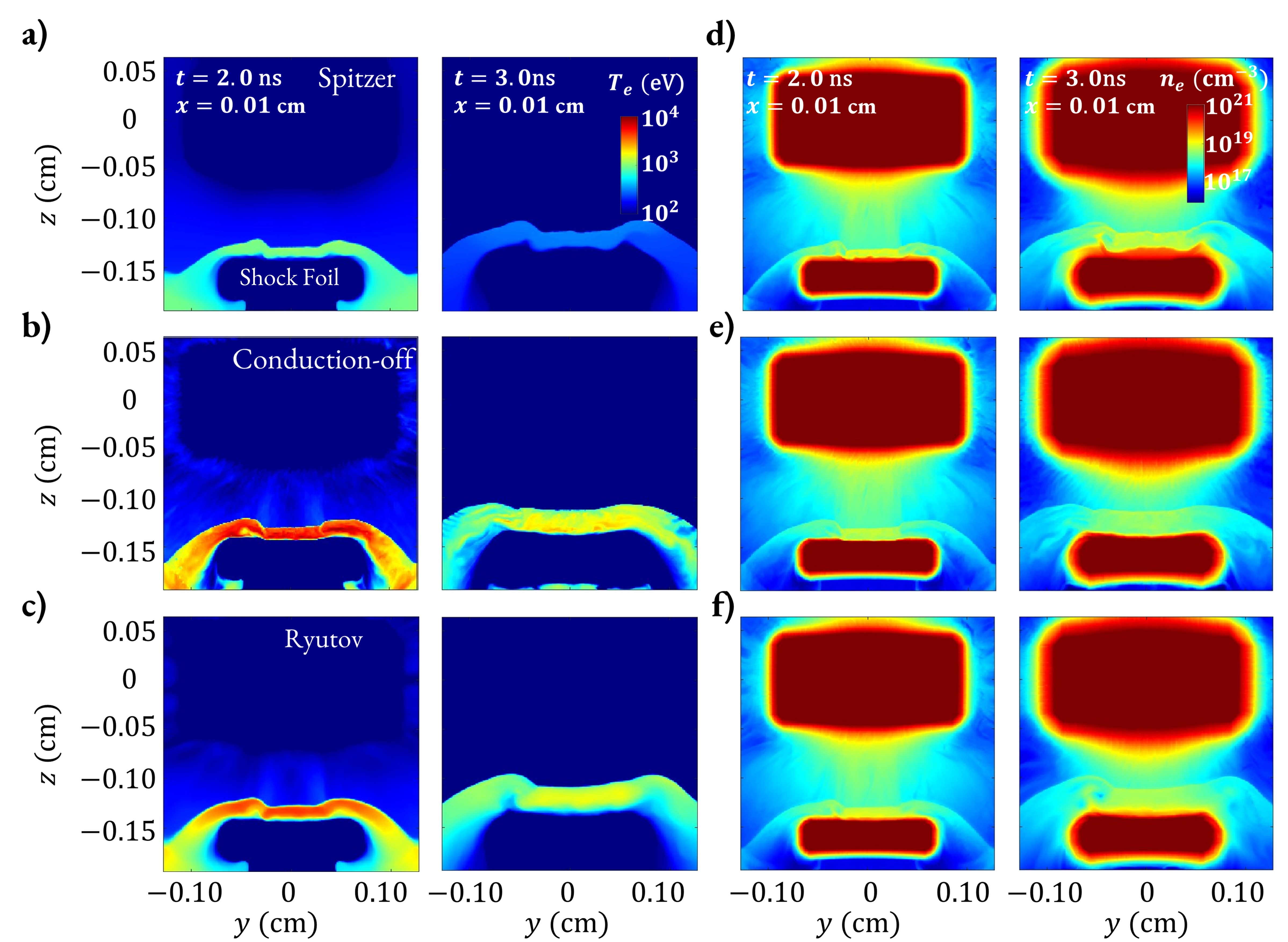}
    \caption{2D visualisations using FLASH simulation data, viewing front-on, into to the drive foil plane, for $T_{e}$ (left-most two columns), and $n_{e}$ (right-most two columns) for Spitzer (a) and (d), conduction-off (b) and (e), and Ryutov (c) and (f) simulations between $2.0~\text{ns}\leq t \leq 3.0$ ns. Slice taken at $x = 0.01$ cm from FLASH simulation domain.}
    \label{fig:x_slices}
\end{figure}

\begin{figure}
    \centering
    \includegraphics[width=1\linewidth]{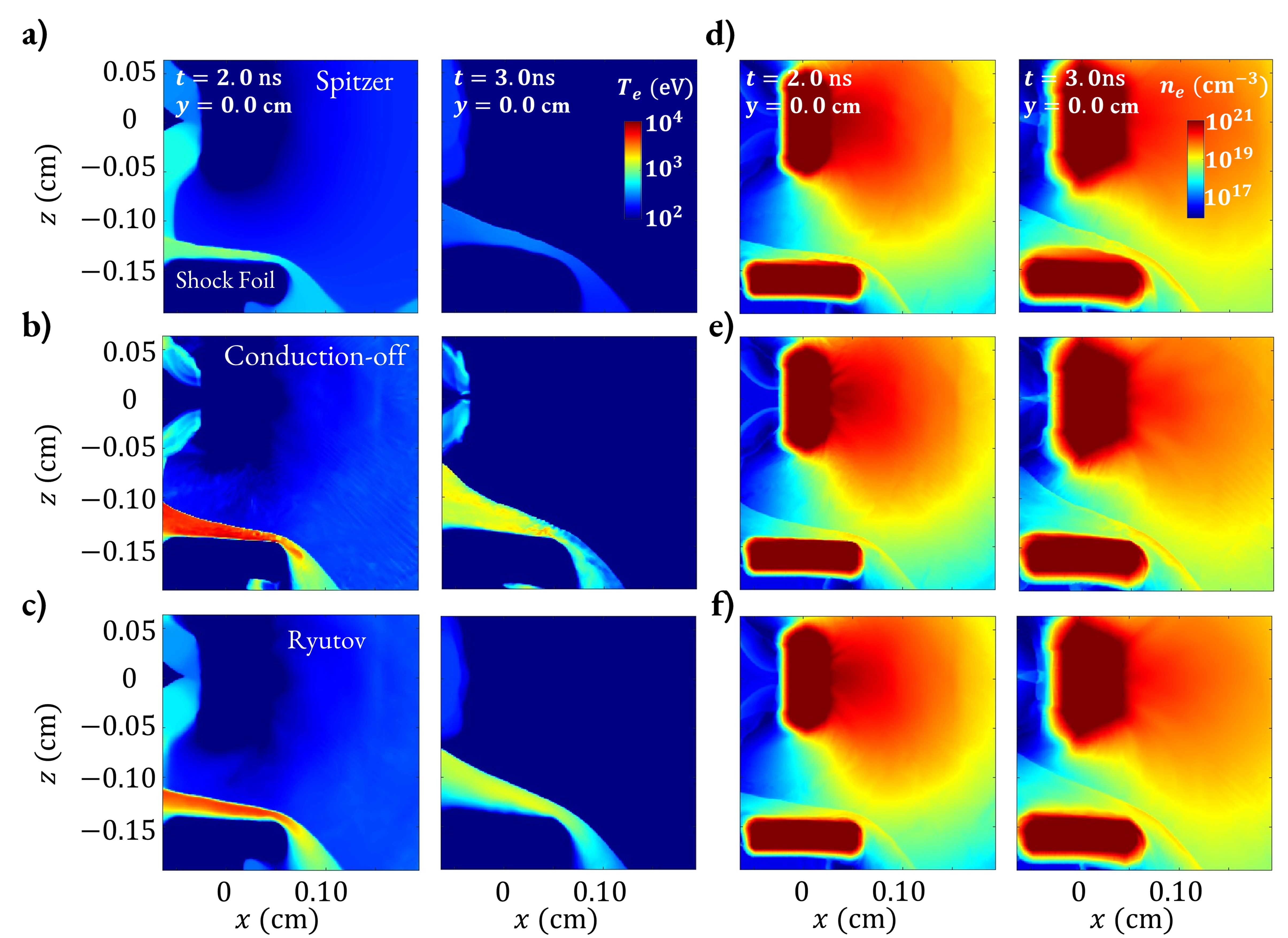}
    \caption{2D visualisations using FLASH simulation data, viewing the side of the drive foil plane normal, for $T_{e}$ (left-most two columns), and $n_{e}$ (right-most two columns) for Spitzer (a) and (d), conduction-off (b) and (e), and Ryutov (c) and (f) simulations between $2.0~\text{ns}\leq t \leq 3.0$ ns. Slices taken at $y = 0.0$ cm from FLASH simulation domain.}
    \label{fig:y_slices}
\end{figure}

The time evolution of the electron temperature $T_{e}$, and electron number density $n_{e}$, for the Spitzer-conduction simulations can be seen in figures \ref{fig:x_slices} and \ref{fig:y_slices} (a) and (d). 
Here, 2D slices have been taken from the simulation domain between times $2.0~\text{ns}\leq t\leq 3.0~\text{ns}$ post laser drive. Slices at fixed $x=0.01~\text{cm}$ in $n_{e}$ show that, as expected, a hot, planar plasma forms just above the shock foil. Within the region itself, a central ($-0.04~\text{cm}\lesssim y \lesssim 0.04~\text{cm}$) high-density region develops, circumscribed by less dense plasma at its edge, which is associated with temperature gradients along $y$. This latter feature is explained by the interaction between the two ablated plasma plumes generated by the LP beams. As they collide, two counter-propagating shock waves develop, with higher-density plasma present between them. This feature appears in figure \ref{fig:x_slices}(d) as a bright column along $z$ at early times that disperses at late times.

The evolution of $T_{e}$ and $n_{e}$ slices between $2.0~\text{ns}\leq t\leq 3.0~\text{ns}$ in the conduction-off simulations is shown in figures \ref{fig:x_slices} and \ref{fig:y_slices}, (b) and (e). The density profile is relatively unchanged in the absence of conduction, although slightly lower densities across the plasma of interest are observed. However, temperatures are nearly an order of magnitude higher when compared with the Spitzer-conduction simulations, and a much more structured temperature profile is observed. Finally, the evolution of $T_{e}$ and $n_{e}$ slices between $2.0~\text{ns}\leq t\leq 3.0~\text{ns}$ with the Ryutov model can be seen in figures \ref{fig:x_slices} and \ref{fig:y_slices}, (c) and (f). The density profile is, again, comparable to those obtained with the other conduction models. Temperatures for the Ryutov-conduction simulation are also much greater than those realised in the Spitzer-conduction simulations, though not as great as those with no conduction at all. At later times, the temperatures seen in the Ryutov-conduction simulation converge towards the conduction-off simulations, although a smoother profile is maintained. This latter finding is likely due to localized thermal conduction, which is absent in the conduction-off simulation. The significant difference in temperatures obtained in these three simulations indicate a dominant dependence of heat conduction on the evolution of the plasma's temperature. 

Beyond qualitative comparisons, in order to quantify characteristic physical parameters in the shocked plasma in each simulation, we carried out the following algorithmic calculation. Firstly, to ensure only the drive foil material was considered, the drive foil material was tagged to allow for easy removal of any shock foil plasma. This helped distinguish between the drive and shock foil post plasma formation. Secondly, only the post-shock plasma is of interest, not the ablation jet, so any plasma that had a negative (i.e., downward) supersonic Mach number, $M_{z} = v_{z}/c_{s}$, was removed, where $v_{z}$ is the plasma velocity along the $z$ direction, and $c_{s}$ is the sound speed in the plasma. Finally, from the remaining plasma, a spatial region within the boundary of the shock foil but in front of the drive foil was chosen such that $B_{y}$ was maximised while also maintaining large $T_{e}$ and $n_{e}$. This was due to $\text{Ha}_{e} > 1$ begin required for the whistler heat-flux instability to grow.

\begin{figure}
    \centering
    \includegraphics[width=1\linewidth]{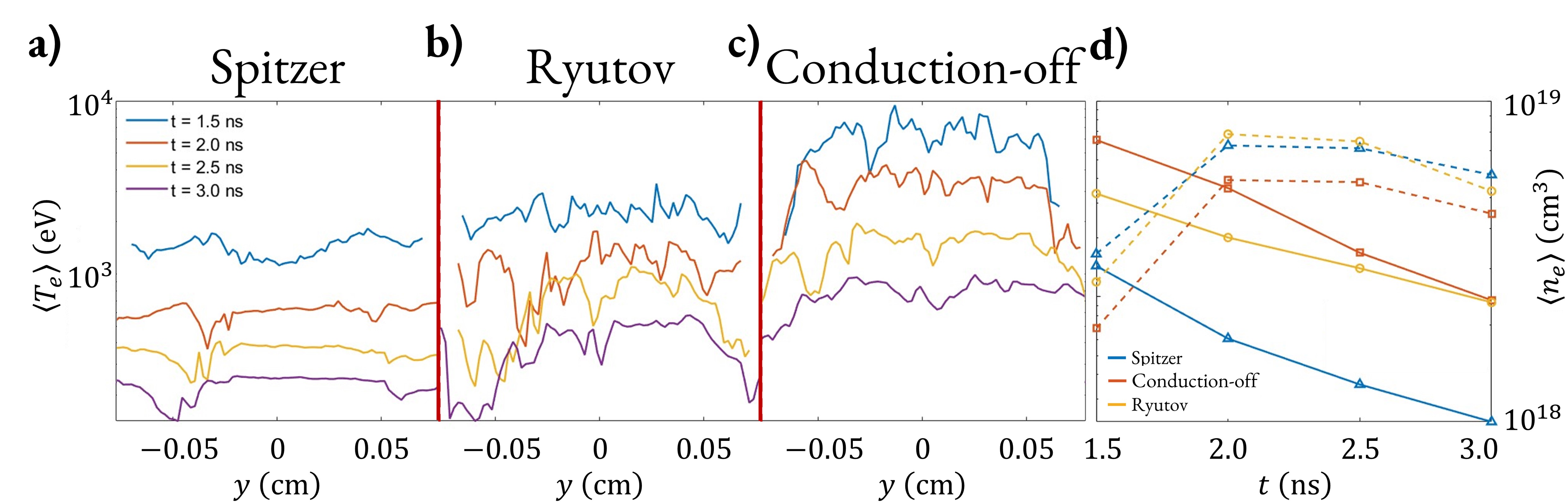}
    \caption{(a) Lineouts of average $T_{e}$ along $y$, post Mach number truncation, integrated in $z$, of the shocked plasma for Spitzer simulation. The blue, orange, yellow, and purple lines corresponding to $t = 1.5,~2.0,~2.5,~\text{and}~3.0~\text{ns}$, respectively. (b-c) same as (a), but for the Ryutov, and conduction-off simulations. (d) Similarly averaged $T_{e}$ (solid line), and $n_{e}$ (dashed line), for Spitzer (blue), Ryutov (yellow), and conduction-off (orange) at each time-step in the shocked plasmas evolution. Averaging  between $-0.07~\text{cm}\leq y \leq 0.07~\text{cm}$ and $x \geq 0.03~\text{cm}$.}
    \label{fig:cond_model_comparisons}
\end{figure}

Averaged electron temperatures and densities, as well as profiles of $T_e$ along the $y$ direction, calculated using this procedure, are shown for each of the conduction models in figure \ref{fig:cond_model_comparisons}. 
In the Spitzer-conduction simulations, at $2.0$ ns after laser drive, the shocked plasma has a $T_{e} \approx 0.61~\text{keV}$, with $n_{e} \approx 8.1\times10^{18}~\text{cm}^{-3}$; in conduction-off simulations, $T_{e} \approx 3.6~\text{keV}$, with $n_{e} \approx 5.7\times10^{18}~\text{cm}^{-3}$; finally, in Ryutov-conduction simulations,  $T_{e} \approx 2.0~\text{keV}$, with $n_{e} \approx 7.8\times10^{18}~\text{cm}^{-3}$, all averaged across the width and front end of the shock foil ($-0.07~\text{cm}\leq y \leq 0.07~\text{cm}$ and $x \geq 0.03~\text{cm}$), and the whole of the $z$ domain above the shock foil. Despite the conduction-off simulations starting with far higher temperatures than the other conduction models it is clear that there is still significant cooling within the plasma. This is likely dominated by adiabatic expansion, and at late times ($t\gtrsim 2.5$ ns), the conduction-off and Ryutov temperature profiles begin to converge. 

\begin{figure}
    \centering
    \includegraphics[width=1\linewidth]{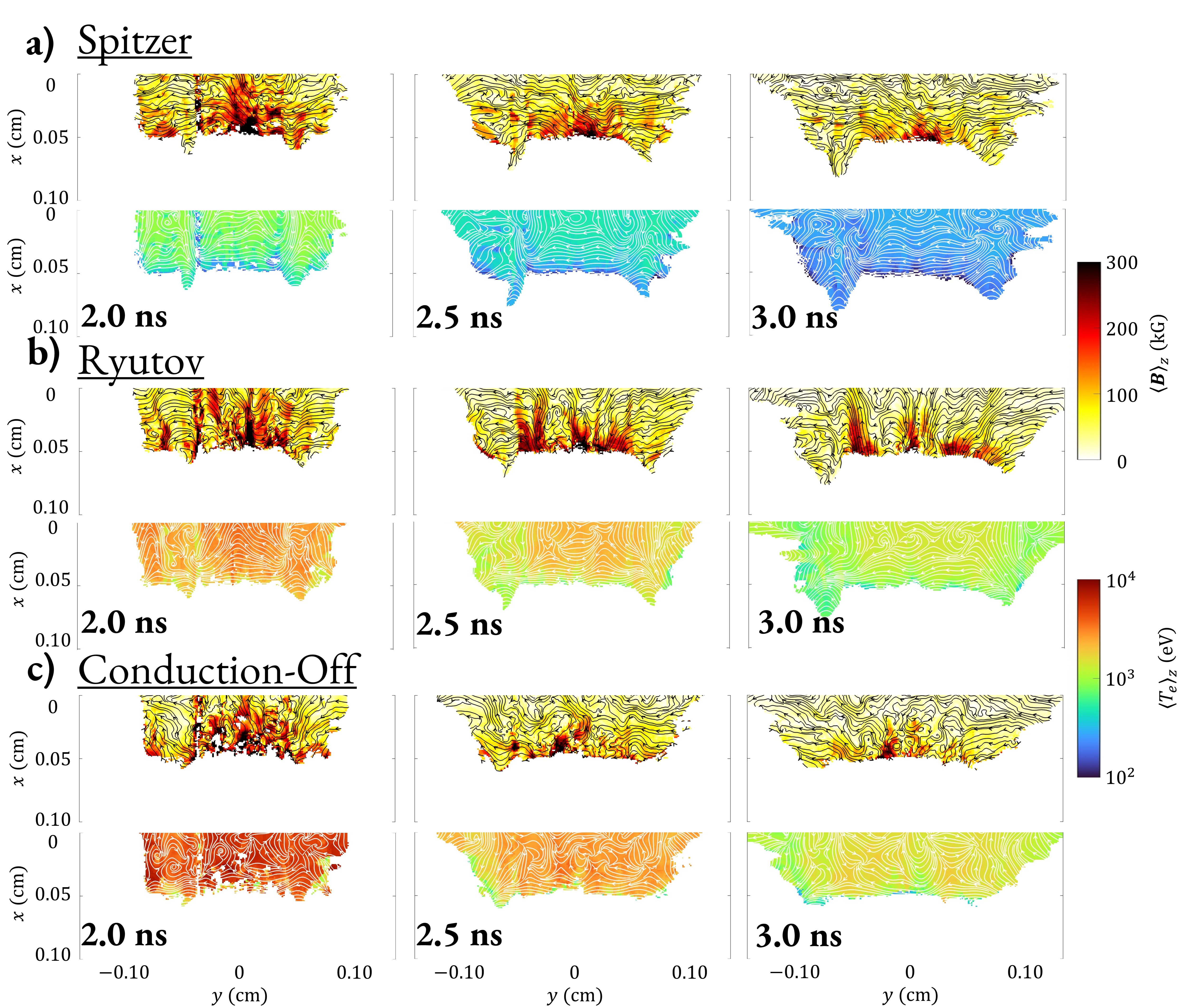}
    \caption{2D visualisations using post-processed FLASH simulation data, viewing top-down, into to the shock foil plane. (a) Top row: Mach number truncated averaged magnetic field morphology in $(x, y)$, integrated across $z$ for the Spitzer-conduction simulations between $2.0~\text{ns}\leq t\leq 3.0~\text{ns}$, with arrows indicating direction of magnetic field. The bottom row shows the electron temperature gradient vector lines overlaid on the temperature maps for the corresponding time, similarly integrated in $z$ across the same range and Mach number truncation. (b-c) Same as (a) but for the Ryutov, and conduction-off simulations respectively.}
    \label{fig:magFieldMorphology}
\end{figure}

The morphology of the magnetic field, along with the temperature gradients for each conduction model evolving with time, is represented in figure \ref{fig:magFieldMorphology}. 
These vector plots were generated by integrating all the variables across the whole 
$z$ domain above the shock foil, post Mach number truncation. To ensure small scale structures $\mathcal{O}(\rho_{e})$ were not affective interpretations, the $T_{e}$ were calculated by applying a $25\times10^{-3}~\text{cm}$ Gaussian smoothing to the FLASH generated $T_{e}$ maps. For the Spitzer-conduction model (figure \ref{fig:magFieldMorphology}(a)) the direction of the magnetic field is predominantly anti-parallel to the y-axis through the plasma of interest, showing the most deviation from this at $2~\text{ns}$ and straightening as time progresses. The direction of temperature gradients remains mostly constant over time, also directed anti-parallel to the y-axis, similarly showing the most deviations from this path earlier in time. The Ryutov-conduction simulations (figure \ref{fig:magFieldMorphology}(b)) exhibit temperature gradient morphology similar to those seen from the Spitzer-conduction simulations, though the magnetic fields have a turbulent nature in the central region. The scale of these fluctuations seem to evolve with time, with the field structures at $3~\text{ns}$ being significantly larger than at $2.5~\text{ns}$. Turbulent structures predominantly grow near the front edge of the shock foil (at $x\approx 0.075~\text{cm}$), while the field lines in the bulk of the plasma remain more straight. Finally, the conduction-off simulations show no preferred direction in the temperature gradients. The magnetic field strength and direction at $2~\text{ns}$ are similar to the Ryutov case, with tightly coiled turbulent structure nearer the edge of the shock foil. At $2.5~\text{ns}$ there is still a favoured direction of the field, anti-parallel to the y-axis, though the field lines begin to break up. By $3~\text{ns}$ the magnetic field lines in the central region have completely broken up, leaving a stochastic magnetic field. Despite the structural variations in the magnetic field between conduction models, the total field strength remain similar, though the differences increase along the $y$-axis for $\mathbf{B}_{y}$. At $2.0~\text{ns}$ after laser drive, in the Spitzer-conduction simulations the shocked plasma has a $|\mathbf{B}| \approx 117~\text{kG}$, with $|\mathbf{B}_{y}|\approx 91.1~\text{kG}$; in Ryutov-conduction simulations, $|\mathbf{B}| \approx 132~\text{kG}$, with $|\mathbf{B}_{y}|\approx 84.8~\text{kG}$; finally, in conduction-off simulations, $|\mathbf{B}| \approx 143~\text{kG}$, with $|\mathbf{B}_{y}|\approx 78.9~\text{kG}$, all averaged across the same region as for the temperature and density estimates, ($-0.07~\text{cm}\leq y \leq 0.07~\text{cm}$ and $x \geq 0.03~\text{cm}$), and the whole of the $z$ domain above the shock foil. 

\begin{figure}
    \centering
    \includegraphics[width=1\linewidth]{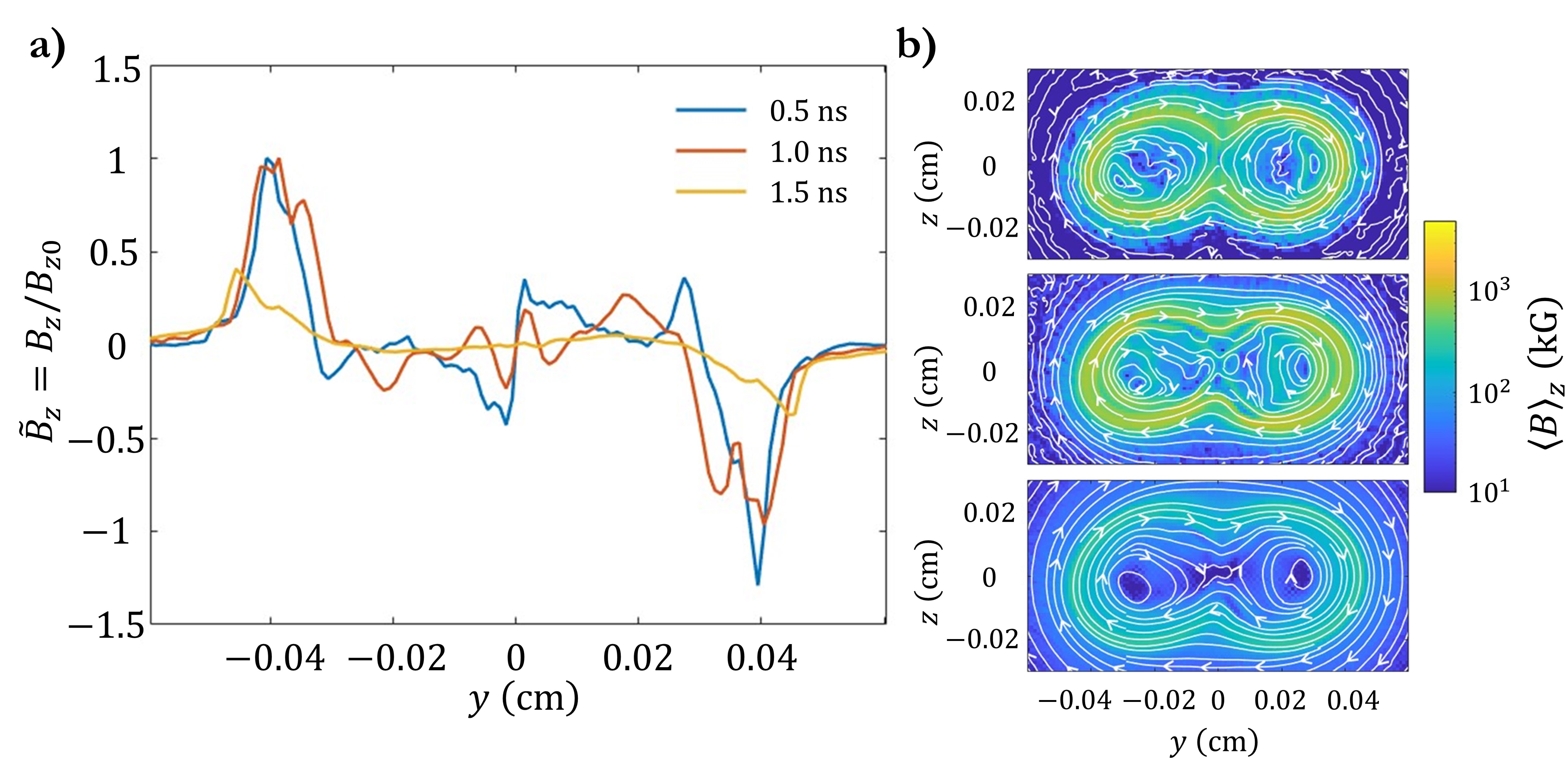}
    \caption{(a) Lineouts taken from $z = 0.0$ cm of slices shown in (b) for the normalised $B_{y}$ component of the field. (b) 2D visualisations using FLASH simulation data of $|\boldsymbol{B}|$ integrated between $0.008~\text{cm} \leq x \leq 0.01$ cm for $t = 0.5,~1.0,~1.5$ ns (top, centre, and bottom respectively) post the drive foils ablation, with arrows indicating the mean field direction.}
    \label{fig:mag_reconnection}
\end{figure}

One subtlety to consider in this experiment is the possible role of magnetic reconnection on both the magnetic-field morphology and the plasma's dynamics. Magnetic reconnection arises naturally when driving a target with two, non-overlapping, powerful laser spots \citep{nilson2006magnetic, fox2011fast, tubman2021observations}. When the front-side blow-off plasma forms, providing they are sufficiently separated, magnetic fields generated via the Biermann battery in each plume can interact. The generated fields will be directed anti-parallel to one another, and when squeezed together, can reconnect. The result of such a change in the magnetic field's geometry is the formation of jets produced orthogonal to the reconnected field due to the magnetic energy being converted into kinetic energy for the charge carriers in the plasma \citep{valenzuela2024x}. Investigating this phenomenon in the Spitzer-conduction simulations, we find that at early times during the ablation $0.5~\text{ns}\lesssim t \lesssim 1.0~\text{ns}$, a small reconnection event has occurred (figure \ref{fig:mag_reconnection}).
However, due to the comparatively small separation of the drive spots compared with the focal spot diameters, temperature and density gradients sufficient to generate large magnetic fields were not induced. As such, insufficient magnetic energy will have been imparted into the ablated plasma to cause any measurable effects on the shocked planar plasma.

\section{Diagnostics} \label{sec:diagnsotics}
In addition to providing physical insights, FLASH simulation data can be used to generate synthetic diagnostic outputs, and thereby improving the design of this experimental platform. These outputs have primarily been used to check the feasibility of fielding particular diagnostics and to aid in their optimal configuration on the Orion Laser Facility. In particular, these synthetic outputs provide useful insights into how differences between conduction models can be measured.

We propose using X-ray spectroscopy to infer $T_{e}$ via the evolution of the spectral chlorine (Cl) lines within the shocked plasma, in particular the relative line intensity combinations from the K-shell, L-shell emission lines and dielectronic satellites for qualitative analysis, and spectral fitting for final $T_{e}$ estimates. Our use of chlorine-doped plastic for the drive foil, but not in the shock foil, is primarily to aid in distinguishing any emission from the drive-foil plasma with that of the shock foil. Another benefit of using chlorine is that the plasma produced will be optically thin to the emission lines of interest across the whole density and temperature space that it occupies. In addition to X-ray spectroscopy, a two-dimensional spatially resolved gated x-ray detector (GXD) will be used to infer the spatial scale of temperature variation within the shocked plasma, as well as corroborating the x-ray spectroscopy measurements. This measurement will be obtained from a pair of spatially resolved filtered self-emission images of the plasma of interest, and then using spectral collisional-radiative codes to relate their ratio to a particular electron temperature. Magnetic field strengths will be inferred using proton imaging, via protons accelerated from SP-irradiated Au foils by target-normal-sheath-acceleration (TNSA). Finally, a combination of spectral line and continuum analysis, alongside the analysis of caustic broadening within proton images~\citep{bott2021inefficient} will be utilised to infer $n_{e}$. To generate the two simulated x-ray diagnostics outputs, we run the collisional-radiative spectral analysis code SPECT3D \citep{macfarlane2007spect3d}, using opacity and EoS tables from PrOpacEoS, and assume a collisional-radiative (non-LTE) plasma, with the FLASH simulation data as input for the spatially resolved plasma parameters. The open-source Python 3 library PlasmaPy \citep{plasmapyCommunity_2024} has been used to generate the synthetic proton radiographs, using the simulated magnetic-field  data from FLASH.

\subsection{Gated x-ray detector diagnostic} \label{sec:GXD}

X-ray self-emission of the shocked plasma has significant contributions from both free-free bremsstrahlung, and also line emission. The plasma of interest is optically thin to these emitted x-rays and as such, the intensity $I$ incident at the virtual detector will have the following relation: 
\begin{equation}
I \propto \int\text{d}\omega \, \hat{R}(\omega)\int\text{d}s~\varepsilon_{\text{plas}}(n_e(s),T_e(s),\omega) \; , 
\end{equation}
where $s$ is the path length of the x-ray radiation, $\omega$ is its frequency, 
\begin{equation}
\varepsilon_{\text{plas}} = \mathcal{A}_{\rm br} n_{e}^{2}T_{e}^{-1/2}\exp\left(-\frac{\hbar\omega}{k_{B}T_{e}}\right) + n_{e} {g}_{\rm line}(T_{e}) 
\end{equation}
 is the emissivity of the plasma accounting for bremsstrahlung and line emission~\citep{rybicki1991radiative}, with $\mathcal{A}_{\text{br}}$ a collection of constants attributed to bremsstrahlung emission, $g_{\text{line}}(T_{e})$ the temperature dependence of the line emission, and $\hat{R}(\omega)$ the frequency-dependent transmission function of a filter material convolved with a microchannel plate (MCP) response function \citep{rochau2006energy} that imitates that of the camera to be used during the campaign. The emission-weighted temperature of any plasma can be inferred by producing two simultaneous images with distinct filters and computing their ratio,
\begin{equation}
  \mathcal{R}(T_{e}) = \frac{ \int\text{d}\omega \, \hat{R}_1(\omega)\int\text{d}s~\varepsilon_{\text{plas}}(n_e(s),T_e(s),\omega) }{ \int\text{d}\omega \, \hat{R}_2(\omega)\int\text{d}s~\varepsilon_{\text{plas}}(n_e(s),T_e(s),\omega) },
\end{equation}
as a consequence of the frequency dependence of filter transmission functions and the temperature dependence of the plasma's emissivity, provided the filters are appropriately selected. From this, a spatially resolved map of the emission-weighted temperature can be generated \citep{poole2026Git}. 
 The emissivity functions through each filter can be obtained through a series of single cell simulations across the range of temperatures, $10~\text{eV}\leq T_{e}\leq 5~\text{keV}$, and densities, $10^{18}~\text{cm}^{-3}\leq n_{e}\leq 10^{19}~\text{cm}^{-3}$, that are expected to be present in the plasma across the times simulated. Therefore, a filtered x-ray ratio curve can be produced to infer a spatially resolved line-of-sight integrated, emission-weighted, temperature profile for the observed plasma. This method of inferring temperature has been used to validate several FLASH simulations against experimental data, as well as being benchmarked against alternate methods for inferring temperature in laser produced plasma such as optical Thomson scattering (OTS) \citep{bott2021time, meinecke2022strong, poole2024Development}.

\begin{figure}
    \centering
    \includegraphics[width=1\linewidth]{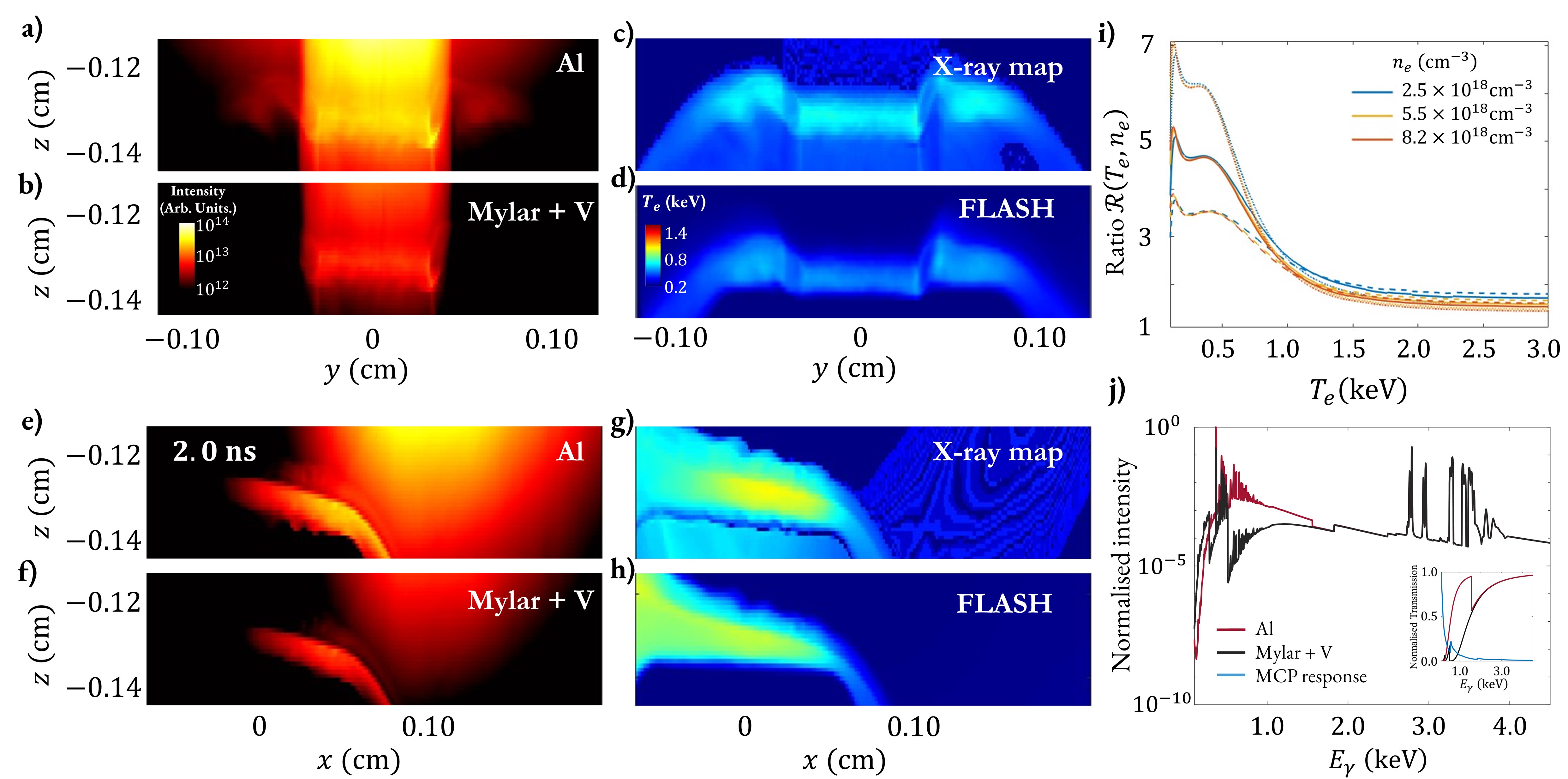}
    \caption{(a) Rear view of synthetic x-ray self emission of shocked plasma at $t=2.0~\text{ns}$ filtered through $0.5~\mu$m Al, convolved with a detector response transmission curve (see insert of (j)). (b) Same as (a), but filtered with $0.9~\mu$m Mylar and $0.4~\mu$m V. (c) Filtered x-ray ratio to temperature map, using ratio curve from (i) at fixed $n_{e} = 6\times10^{18}~\text{cm}^{-3}$. (d) LoS, mass weighted, averaged $T_{e}$ from FLASH simulation data at $t=2.0~\text{ns}$. (e-h) Same as in (a-d) but for the side view of the plasma. (i) Filtered x-ray to temperature ratio curve at a range of fixed densities between $2.5\times10^{18}~\text{cm}^{-3}\leq n_{e}\leq 8.2\times10^{18}~\text{cm}^{-3}$. The dashed and dotted lines correspond with the $+20\%$ and $-20\%$ thickness errors on the filter materials respectively. (j) Single-cell spectra at $T_{e} = 1,542~\text{eV}$, and $n_{e} = 5.5\times 10^{18}~\text{cm}^{-3}$ attenuated by an Al filter in red, and Mylar + V in black. Both filter transmission functions have been convolved with an MCP response function. The insert shows the frequency dependent filter transmission functions, with the MCP response in blue.}
    \label{fig:ratioCurve_and_transFilter}
\end{figure}

The filter combinations selected for this experimental campaign are $0.5~\mathit{\mu}\text{m}$ of aluminium (Al), and $0.9~\mathit{\mu}\text{m}$ of Mylar with $0.4~\mathit{\mu} \text{m}$ of vanadium (V). 
The frequency-dependent transmission functions for the Al and Mylar/V filters, and the detector response, are shown in figure \ref{fig:ratioCurve_and_transFilter}j.
To determine the ratio curve for this filter combination, the frequency-dependent radiative emission of parylene-C was computed using SPECT3D over a wide range of temperatures and densities. These simulations were run assuming quasi-steady state non-LTE for the kinetics model, with PrOpacEOS being used for the EoS data for the parylene-C. The resulting emission spectra were then convolved with $\hat{R}_{\rm Al}(\omega)$ and $\hat{R}_{\rm Myl + V}(\omega)$, and the ratio curve $\mathcal{R}(T_{e})$ subsequently computed. Figure \ref{fig:ratioCurve_and_transFilter}i shows that this filter combination produces a ratio curve that is reliable between $300~\text{eV}\leq T_{e}\leq 1.5~\text{keV}$, due to its monotonically decreasing profile, and in particular it's insensitivity to density fluctuations if $100~\text{eV}\leq T_{e}\leq 1~\text{keV}$. This covers the range of temperatures predicted by the Spitzer-conduction FLASH simulations. 

\begin{figure}
    \centering
    \includegraphics[width=1\linewidth]{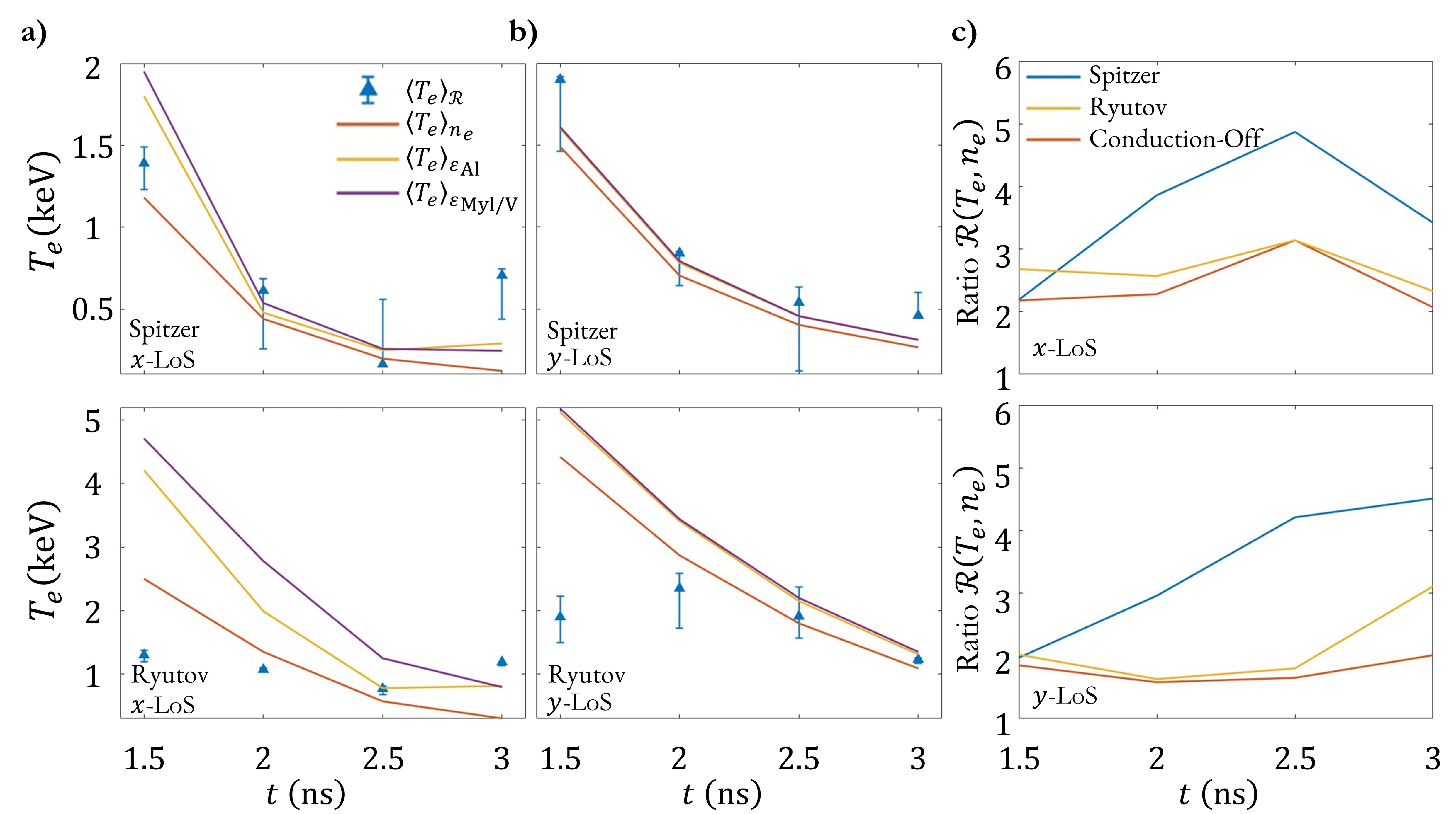}
    \caption{(a) Evolution of LoS integrated, mass weighted, average $T_{e}$ from FLASH compared against the x-ray ratio inferred temperature $\langle T_{e}\rangle_{\mathcal{R}}$ (blue triangles) for $x$-LoS for the Spitzer and Ryutov governed plasma. The orange line is mass weighted, yellow is emissivity weighted as seen through the Al filter, and purple is the same as yellow but for the Mylar and V filter. Reconstructions were performed using ratio curves assuming a $\pm20\%$ thickness error on the filter combinations. The errors on the inferred temperatures were calculated using these thicker and thinner filter ratio curves. (b) Same as (a) but for the $y$-LoS. (c) Evolution of the x-ray filter ratio for each conduction model and LoS. The blue, yellow, and orange lines correspond to the Spitzer, Ryutov, and conduction-off models respectively.}
    \label{fig:AveTempComparison_GXD}
\end{figure}

To test the efficacy of this filter combination for the experimental platform, two virtual 2D spatially resolved GXDs were used to measure the simulated self-emitted soft x-rays from the shocked ablation plasma in the FLASH simulations. 
The synthetic emission was simulated by using the FLASH simulation output as an input to SPECT3D, and then adopting the same methodology as was used for the single-cell ratio curve simulations, but also accounting for opacity and LoS integrating. 
These SPECT3D simulations were run with the shock foil removed; limitations in the FLASH shock modelling, as discussed in section \ref{sec:sim_plasma_param}, and the use of isotropic heat conduction, rather than anisotropic conduction constrained by magnetic fields parallel to the foil, lead to excessive and non-physical heating of the shock foil. Consequently, SPECT3D simulations including the shock foil would confound the analysis and were therefore excluded.
Figure \ref{fig:ratioCurve_and_transFilter} shows synthetic self-emitted x-rays, as seen through both filter combinations and each virtual detector's line-of-sight (LoS), for the Spitzer-conduction simulations at $2.0~\text{ns}$ after the laser drive, along with the ratio curve and the resulting temperature map. The reconstructed temperature map matches well qualitatively with the LoS-integrated mass-weighted temperature calculated directly from FLASH, see figure \ref{fig:AveTempComparison_GXD}.

\begin{figure}
    \centering
    \includegraphics[width=1\linewidth]{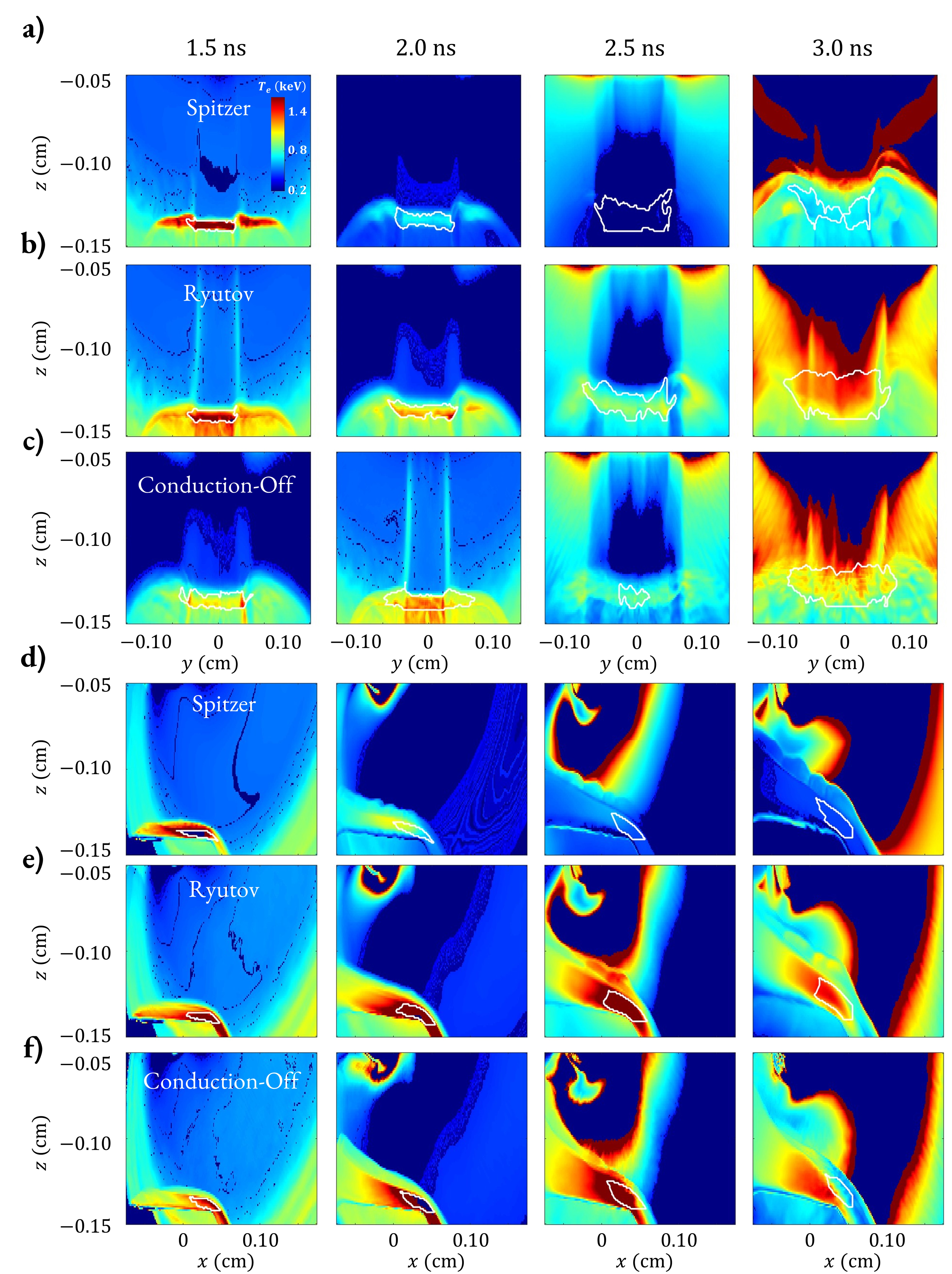}
    \caption{Filtered x-ray ratio temperature maps between $1.5~\text{ns}\leq t\leq3.0~\text{ns}$ for Spitzer (a), Ryutov (b), and conduction-off (c) simulations along the $x$-LoS. (d-f) Same as in (a-c) but for the $y$-LoS. The white outline corresponds to the region selected from the intensity distribution of the shocked plasma, and the region averaged over when comparing $T_{e}$.}
    \label{fig:condModeL_comparisons_GXDMap}
\end{figure}

Having validated the approach in one case, we now apply it to simulations with each conduction model discussed here, over a range of times in the simulations. Figure \ref{fig:condModeL_comparisons_GXDMap} shows the evolution of the inferred temperature maps. 
Qualitatively, the simulated diagnostic captures many key features of the temperature profiles for each of the conduction models, particularly during the first 1 ns after the collision of the front-side blow-off plasma jet with the shock foil. The shocked plasma is observed to be significantly hotter than the pre-shocked jet, and a temperature gradient in the plane of the shock foil is also observed. The simulated diagnostic is also able to distinguish between the conduction models successfully: by 2.5 ns, the inferred temperature from the Spitzer-conduction FLASH simulations is well below that of the other conduction model.

That being said, the diagnostic does not always compute the mass-weighted temperature of the plasma of interest with high fidelity. For all conduction models, the $x$-LoS reconstructions become unreliable after $2.5~\text{ns}$. This is due to higher-density front-side blow-off plasma, which is comparatively emissive to the shocked plasma yet significantly cooler ($T_{e}$$\sim$$ 100~\text{eV}$) overlapping with the line of sight to the shocked plasma. There are also profile shadows observed around the shocked region from the $x$-LoS for both the temperature maps and the LoS integrated mass-weighted averages from FLASH at $t\gtrsim2.5$ ns. Similarly these arise due to the shocked profile having a spatial gradient in $\boldsymbol{\hat{z}}$, resulting in emission from the top of the shocked region at different positions along the camera's LoS. This effect can also be seen in the $y$-LoS reconstructions. As the time increases, contributions from the jet start to appear within the viewing window and begin to overlap with the front-side of the shocked plasma along the $x$-LoS. This apparent temperature increase of the blow-off plume comes from the structure of the ratio curve. For $10\lesssim T_{e}\lesssim100~\text{eV}$, the profile has a flip in average gradient sign, meaning that a decreasing filtered emission ratio corresponds to a decrease in temperature. The result of this is that across the entire temperature domain, inclusive of plume and shocked plasma, there does not exist a one-to-one map between the filtered emission ratio and temperature for the filters chosen here.
The $y$-LoS images of the plasma is less affected by the emission from the blow-off plume, though it exhibits a high temperature shell around the interaction region due to the limitations of the ratio curve discussed above. Based on this analysis, we conclude that the most reliable LoS integrated, mass weighted temperature is likely to be inferred from the $y$-LoS maps, though the $x$-LoS is still of use as an indicator of temperature gradients for $t\leq 2.5~\text{ns}$. 

\begin{figure}
    \centering
    \includegraphics[width=1\linewidth]{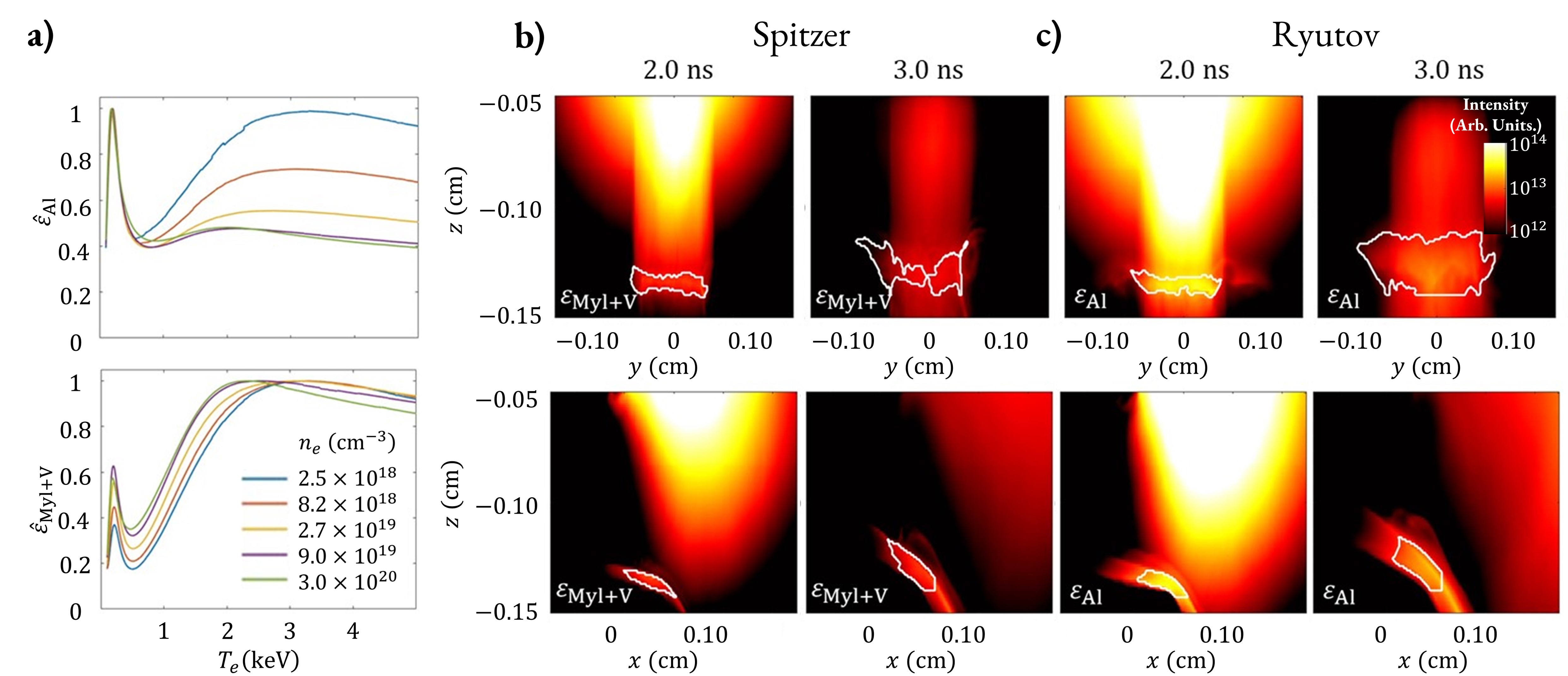}
    \caption{(a) Single-cell generated, frequency integrated, plasma emissivity as seen through $0.5~\mathit{\mu}\text{m}$ Al (top), and $0.9~\mu$m Mylar and $0.4~\mu$m V (bottom) filters, and transformed by an MCP response function. Both sets of curves have been normalised to their respective maximum value uniquely for each density. (b-c) Synthetic filtered self-emitted x-rays as seen through $x$-LoS (top row), and $y$-LoS (bottom row) over time for Spitzer, and Ryutov governed plasma, respectively. The white outline corresponds to the region selected from the intensity distribution of the shocked plasma.}
    \label{fig:emissivityDependencies}
\end{figure}

 For quantitative analysis, we average inferred temperatures over regions selected systematically from the synthetic x-ray images. The selection process involved computing 
 contours of the emission detected from the shocked plasma, then retaining all data from regions emitted at least 33\% of its peak. Though the resulting inferred temperatures were insensitive to the specific truncation percentage, the value to truncate was chosen to maximise the size of the plasma of interest, whilst minimising emission from the background. For temperatures $300\lesssim T_{e}\lesssim1,000~\text{eV}$ and densities $5\times10^{18}\lesssim n_{e}\lesssim10^{19}~\text{cm}^{-3}$ the plasmas' emissivity, as detected by the Al filter and MCP, is more sensitive to temperature fluctuations than to density fluctuations, relative to the emissivity as seen through the Mylar and V filter (figure \ref{fig:emissivityDependencies}a). 
 For temperatures $1,000\lesssim T_{e}\lesssim3,000~\text{eV}$ and densities $10^{18}\lesssim n_{e}\lesssim10^{19}~\text{cm}^{-3}$ the inverse is true. Due to this, for the Spitzer-conduction simulations at $2.0\leq t\leq 3.0~\text{ns}$ the region of interest was selected from the Mylar/V-filtered synthetic x-rays. For all other times and for alternative conduction models, the region was selected using the Al-filtered synthetic x-rays. 

The method for selecting the region of interest for quantitative analysis was slightly more involved for the $x$-LoS images, particularly for $t\geq 2.0~\text{ns}$, due to significant emission from the front-side blow-off plasma jet in front of the shocked plasma. This highly emissive region appears in the centre of the $x$-LoS emission due to the collision of the plumes of plasma during the initial ablation of the drive foil, discussed in section \ref{sec:target_design}. The emission from the jet can be distinguished from the shocked plasma at all times by fitting the former's profile away from the shocked plasma, which is found to be smooth, and interpolating the result over the shocked plasma. The emission profile of the front-side blow-off plasma is determined by computing a series of one-dimensional profiles of the emission in $z$ spanning the entire $y$-domain, with each profile in $z$ starting at z = -0.05 cm (i.e., at the bottom of the drive foil), and terminating when the gradient of that line-out changes sign (i.e., at the beginning of the region of emission coming from the the shocked plasma). An exponential curve was fitted to this emission profile, and then extended into the shocked region. The fitted profile was subtracted from the image,  removing the contribution from the front-side blow-off plasma and revealing the plasma of interest. From this point, the same selection process that used for the $y$-LoS images was adopted, except the bottom one-tenth of the peak emission was removed instead. 

A time-series plot for the inferred temperatures of these selected regions, for each conduction model and LoS, can be seen in figure \ref{fig:AveTempComparison_GXD}.
 For the Spitzer-conduction simulations, the temperature reconstructions agree well with the mass- and emission-weighted values computed directly from FLASH for the $y$-LoS at all times. For the $x$-LoS estimates, there is also good agreement initally, but the reconstructed temperature at $t=3.0~\text{ns}$ erroneously increases. This is most plausibly attributed to contributions from the front-side blow-off plasma. For the Ryutov-conduction simulations, the reconstructed temperature significantly underpredicts FLASH at $t\leq2.0~\text{ns}$ for both lines of sight, before coming into better agreement at later times. The disagreement at earlier times is not surprising, given that the ratio curve is insensitive to temperatures at $T_{e}\sim 3$-$5~\text{keV}$, i.e. the shocked plasma's temperature as predicted by FLASH at these times. 
 Though these reconstructions fail to accurately predict the earlier-time temperatures for the alternative conduction models, it is clear that there is a significant measurable difference between the evolution of the temperature profiles when the plasma is governed by Spitzer compared to models with suppressed conduction. It should be noted that emissivity weighted temperatures were not possible to obtain for the conduction-off model due to the temperature grid not reaching above $5~\text{keV}$; similarly the photon energy grid was such that a non-negligible contribution of emission at frequencies not accounted for in the single-cell simulations -- and hence the emissivity maps -- is likely. As such, the inferred ratio value evolution is shown in figure \ref{fig:AveTempComparison_GXD}c, but not the comparisons with the emissivity weighted temperatures inferred directly from FLASH.

\subsection{X-ray spectroscopy}\label{sec:spectroscopy}
\begin{figure}
    \centering
    \includegraphics[width=1\linewidth]{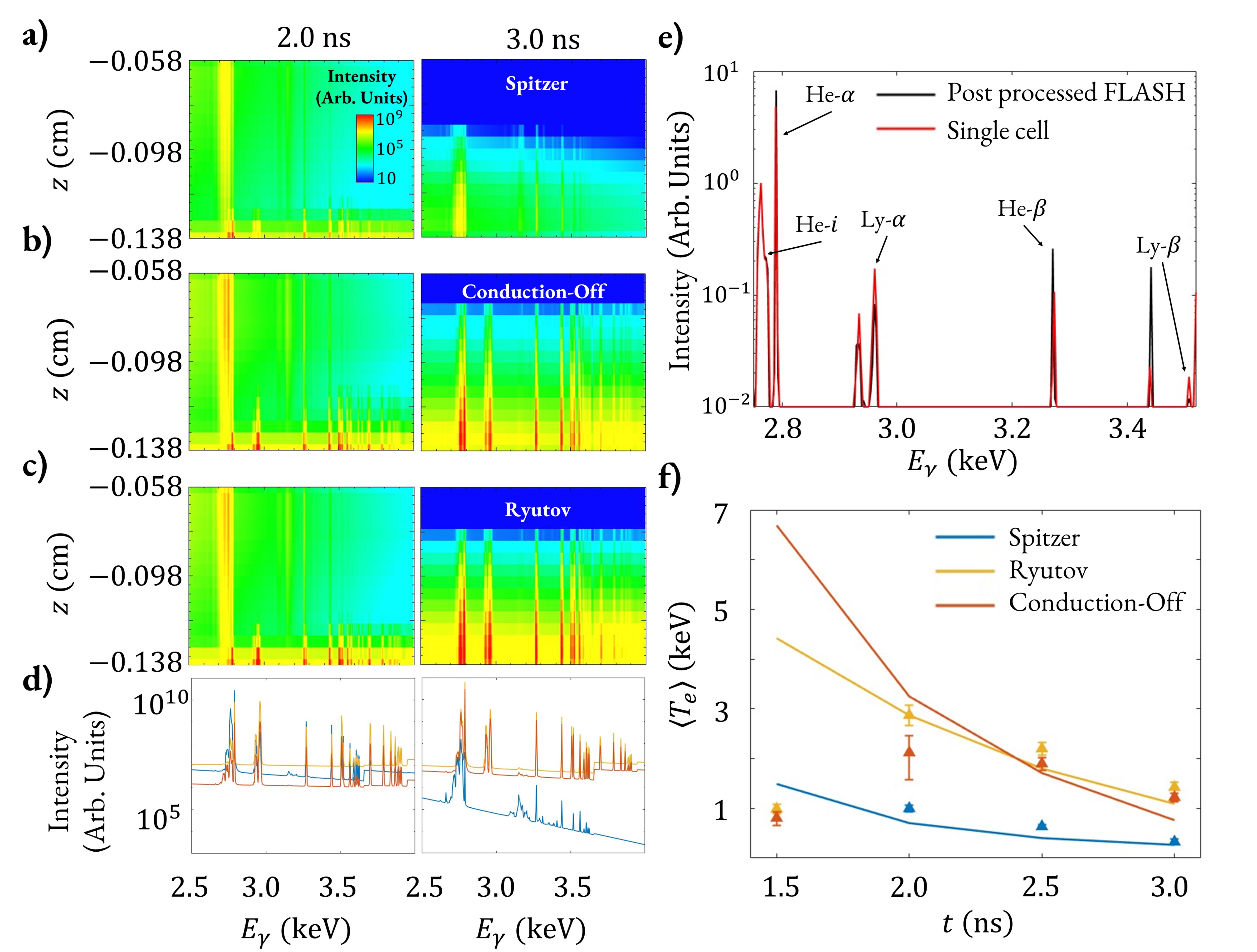}
    \caption{Spatially resolved, synthetic, x-ray spectrographs between $2.0~\text{ns} \leq t \leq 3.0~\text{ns}$ for Spitzer (a), Ryutov (b), and conduction-off (c) governed plasma. (d) Lineouts of spectra for each conduction model taken along the brightest pixel in $z$, with Spitzer, Ryutov, and conduction-off shown in the blue, yellow, and orange respectively. (e) Lineout of spectra from Spitzer governed plasma at $t=2.0~\text{ns}$ in black overlaid with best-fit single cell spectra determined by L2 norm error minimisation in red, with notable lines of interest labelled. (f) Inferred temperatures using best-fit spectra for each conduction model over time shown by the triangles. The solid lines show the mass weighted, LoS integrated $T_{e}$ averages directly from FLASH.}
    \label{fig:condModeL_comparisons}
\end{figure}

Synthetic self-emission x-ray spectra have been generated, using the FLASH simulation data, with the shock foil removed as discussed in section \ref{sec:GXD}, as inputs to SPECT3D, using PrOpacEOS equation of state, and opacity tables. A collisional-radiative (non-LTE) kinetics model was used for the generation of all generated spectra. A virtual spatially resolved, along $z$, x-ray spectrometer with $50~\mu$m spatial resolution has been used to predict the continuum and line self-emission at different times in the evolution of the post-shocked plasma. The viewing window of the virtual detector was $0.8~\text{mm}\times~0.3~\text{mm}$, centred with the coordinate axis matching the FLASH simulation domain. The virtual detector position, and viewing window emulate the spectrometer slit (see figure \ref{fig:targetSchematic}). A variable spectral resolution has been used, increasing grid points between $2.5~\text{keV}\leq E_{\gamma}\leq 4.0~\text{keV}$ as this range contains the spectral lines of interest, where $E_{\gamma}$ is the emitted photon energy. The generated spectrographs have been convolved with a transmission function corresponding to a $25~\mu$m Be sheet which will be a filter used during the Orion campaign for camera shielding purposes. Low energy emissions, $E_{\gamma} \lesssim 1.0~\text{keV}$, will have a transmission fraction of $\lesssim 0.1$ hence, they will be absorbed by the Be filter. As such, many of the H and C lines from the spectrographs will be heavily screened, while Cl lines will be minimally attenuated. Figure \ref{fig:condModeL_comparisons} contains the generated spectrographs between $2.0~\text{ns}\leq t \leq 3.0~\text{ns}$, for all conduction models, along with lineouts taken through the brightest region from the shocked plasmas emission.

\begin{figure}
    \centering
    \includegraphics[width=1\linewidth]{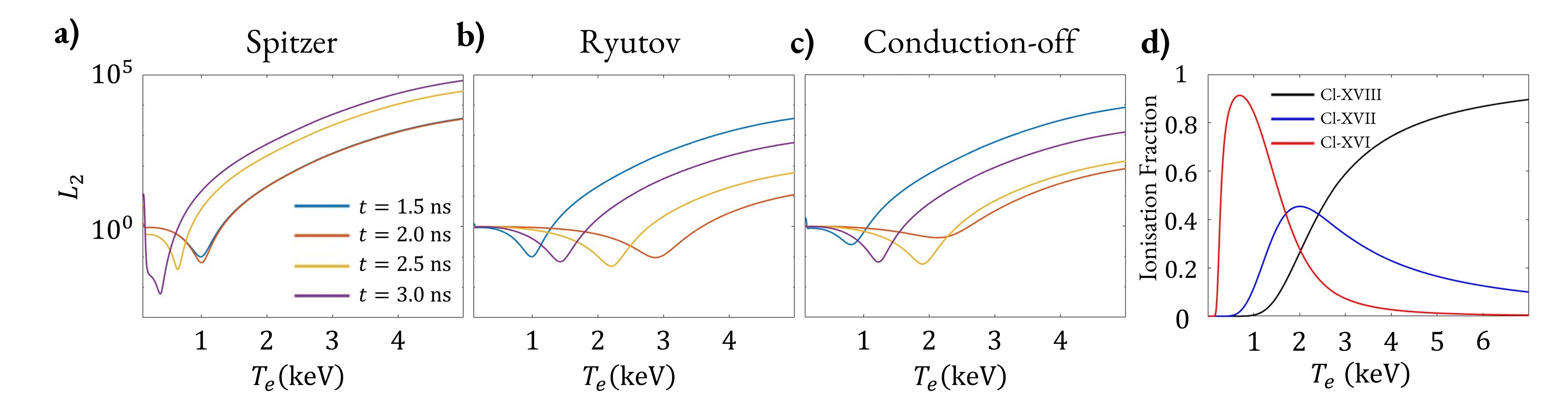}
    \caption{(a-c) $L_{2}$ norm against $T_{e}$ for each conduction model between $1.5~\text{ns}\leq t\leq3.0~\text{ns}$. (d) Ionisation fraction for fully ionised, hydrogenic, and He-like Cl against $T_{e}$ at $n_{e}=6\times10^{18}~\text{cm}^{-3}$.}
    \label{fig:ionRatio}
\end{figure}

Over the investigated timescale, both the shocked ($T_{e} \gtrsim 250$ eV) and jet ($T_{e} \gtrsim 50$ eV) plasma remain near fully ionised, with only the Cl retaining some bound electrons in the shocked plasma, and a small percentage of C and Cl retaining bound electrons in the jet plasma, see figure \ref{fig:ionRatio} for ionisation stage fractions. As a result, a dominant contribution to the x-ray self-emission continuum will be free-free bremsstrahlung radiation. Assuming a thermal distribution of particles, the spectral density can be described by $\epsilon^{\text{ff}}_{\omega} \propto Z_{\text{eff}}n_{e}^{2}T_{e}^{-1/2}\exp{\left(  -\hbar\omega/k_{B}T_{e} \right)}$ \citep{rybicki1991radiative}, where $Z_{\text{eff}} \approx (0.5Z_{\text{H}} + 0.4375Z_{\text{C}} + 0.0625Z_{\text{Cl}})$ is the effective ionisation state of the plasma (with $Z_{\text{H}}$, $Z_{\text{C}}$, and $Z_{\text{Cl}}$ being the charges of H, C, and Cl, respectively), $\omega$ is the photon frequency. With this in mind, the evolution of the plasma can be qualitatively characterised by several indicators from the synthetic spectrographs, namely that it is cooling and that the shocked region is growing. The former can be inferred by the loss of the hydrogenic lines at late times, notably the Ly-$\alpha$, $E_{\gamma}\approx 2961$ eV transition in Cl. This indicates that the population of Cl-XVII has decreased due to cooling, allowing an increased rate of recapture between these hydrogenic Cl ions and the thermal electron population. The latter can be inferred by the vertically growing bright strip in $E_{\gamma}$. At $t = 2.0~\text{ns}$ this strip can be seen to cover the bottom two pixels of the virtual detector image. By $t = 3.0~\text{ns}$ it has grown to a region covering at least five pixels. The background jet emission can be seen to decrease over time, for the energy range of interest. This can be attributed to the jet being cooler, while comparatively dense relative to the shocked plasma. At large $\hbar\omega/k_{B}T_{e}$, $\epsilon^{\text{ff}}_{\omega}$ becomes more sensitive to changes in $T_{e}$ than in $n_{e}$, and as such the emission intensity experiences an exponential cut-off. That is to say, despite the jet and shocked plasma densities converging at late time, the self-emission is decreasing due to a temperature drop. This leads to two distinct domains, namely a hot shocked region surrounded by a high density, cooler background, as is seen in the FLASH outputs.

More quantitatively, performing a fitting procedure to the spectra can give a temperature estimation of the plasma. The region of the spectrum where the Ly-$\alpha$, Ly-$\beta$, He-$\beta$ and He intercombination (He-$i$) line are present was selected for the fitting, $2.7\leq E_{\gamma}\leq 3.5~\text{keV}$. The plasma is optically thin to these lines and they are relatively abundant and the resonance lines are insensitive to changes in $n_{e}$ across the temperature ranges of interest, $T_{e}\gtrsim 300$ eV being ideal for comparison of conduction models. Taking ratios of these K-shell and L-shell transition lines has historically been a measure of temperature, and is a well posed method in the literature for monoelectronic and dielectronic ions \citep{gabriel1972dielectronic, presnyakov1976x}. Each spectrum, single cell and FLASH, was normalised to the He-$\alpha$ line as it appears across all times for each conduction model. As we were interested in fitting to the spectral lines and not the background, anything below $1\%$ of the peak He-$\alpha$ line intensity was set as a flat constant background. From this an $L_{2}$-norm error minimisation was used to estimate, for each conduction model at each time step at fixed density spanning the whole temperature space, the best fit single cell spectra for the post-processed FLASH generated spectrum using the following expression:
\begin{equation}
    L_{2} = \frac{ \int_{\omega_{\text{min}}}^{\omega_{\text{max}}} |f_{\text{SC}}(\omega) - f_{\text{FLASH}}(\omega)|^{2}\text{d}\omega }{\int_{\omega_{\text{min}}}^{\omega_{\text{max}}} |f_{\text{FLASH}}(\omega)|^{2}\text{d}\omega},
\end{equation}
where $[\omega_{\text{min}},~\omega_{\text{max}}] = [2.7~\text{keV}, ~3.5~\text{keV}]$ are the photon energy boundaries containing the spectral lines of interest, and $f_{\text{SC}}(\omega)$ and $f_{\text{FLASH}}(\omega)$ are the single cell and FLASH spectra respectively. The temperature that corresponded to the best-fit single cell was used as the inferred temperature. The temperature dependant $L_{2}$ functions can be seen in figure \ref{fig:ionRatio}. The uncertainty in the inferred temperatures was calculated by finding the $T_{e}$ boundaries of $1.5\times\min{[L_{2}(T_{e})]}$, that is, $\pm50\%$ of the global minimum for each of the $L_{2}$ curves. For all conduction models there is good agreement between the FLASH averages and the spectrally inferred values, particularly for the Spitzer plasma. The inferred temperatures at $t=1.5~\text{ns}$ for the alternative conduction models are significantly lower than expected compared to the FLASH averages. At the earlier times, particularly for the Ryutov and conduction-off models, the jet plasma has a high enough temperature for the hydrogenic and He-like Cl lines to be present. Hence, these lines contribute to the same spectral features as in the shocked plasma, likely reducing the inferred temperature. In addition, the temperature domain used in the spectral simulation was limited to a maximum of $T_{e}=5$,$000~\text{eV}$. This was selected based on the SPECT3D upper photon energy limit of $E_{\gamma}=30~\text{keV}$, and to avoid the temperature grid becoming too sparse. This cap artificially lowers the maximum temperature of the plasma in the hottest regions. The early-time under-predictions are consistent with the combined influence of these two effects. The FLASH temperatures used for the comparison were from the $y$-LoS averages, as the $x$-LoS are averaged over the space containing significant contributions from the plasma plume. Unlike the GXD analysis, see section \ref{sec:GXD}, which includes contributions from both continuum and spectral line emission, here only spectral lines are used in the fitting. Therefore, the observed Cl lines will originate primarily from the hotter shocked plasma and not from the cooler plume.

The techniques used here illustrate that there is a significant and measurable difference in the evolution of the emission spectra between the the Spitzer and alternative conduction models, both qualitatively and quantitatively. A good example can be seen in the evolution of the Ly-$\alpha$, and its satellite lines. By $t=3.0~\text{ns}$ these lines have vanished from the post-processed Spitzer simulations, but remain relatively unchanged for the Ryutov and conduction-off simulations.

\subsection{Proton Imaging}\label{sec:PRAD}
In this experiment, the MeV protons to be used for imaging magnetic fields are generated by a SP laser. A foil irradiated by a high-intensity laser beam can generate protons with a Maxwellian-like distribution of energies from the rear side via TNSA, which operates as follows. The laser's electric field energises and accelerates electrons through the target. A sheet of electrons is ejected from the rear side of the foil, causing a charge imbalance that in turn accelerates protons \citep{macchi2013ion}. These protons traverse the plasma of interest and interact with the magnetic fields within it via Lorentz forces. These interactions distort the beam profile, imparting the bulk of protons with field structure information in the form of a 2D image detected by a radiochromic film (RCF) stack. Provided that the proton deflection angles due to magnetic fields are of order unity or smaller than the beam's divergence angle, the measured fluence profile can be used to infer a path-integrated magnetic field \citep{kugland2012invited, bott2017proton, schaeffer2023proton}. 

\begin{figure}
    \centering
    \includegraphics[width=1\linewidth]{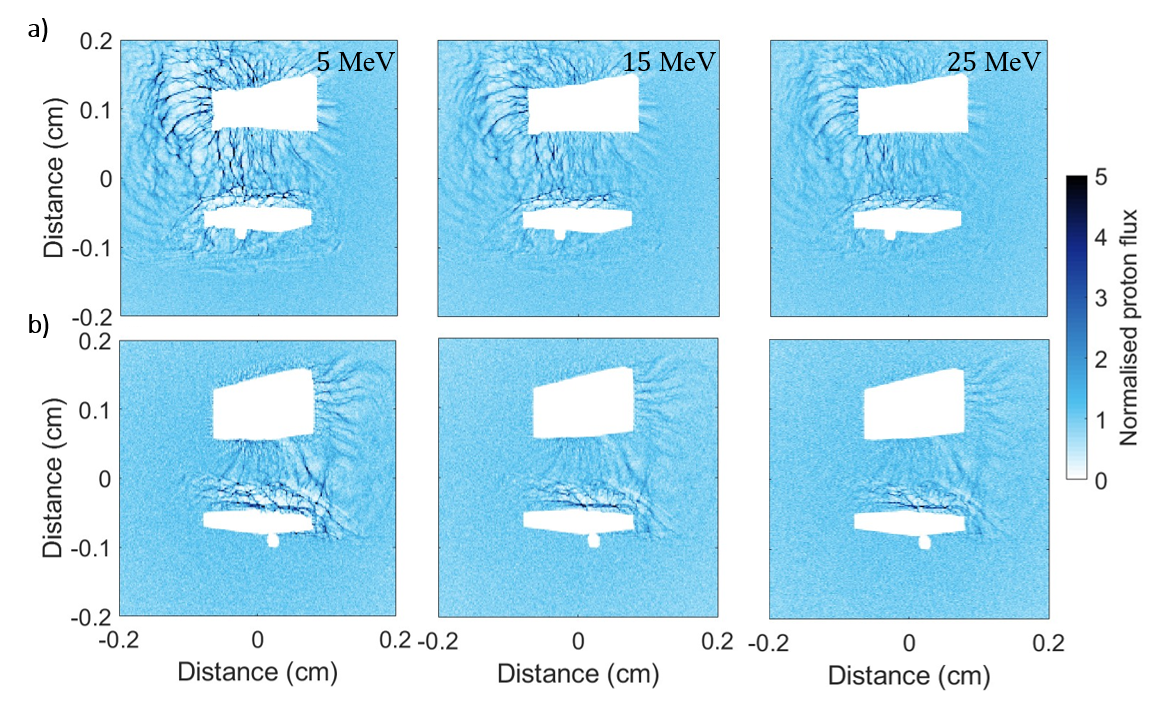}
    \caption{Synthetic proton radiographs for virtual detector positions: (a) view 1 ($60^{\circ}$ clockwise in the $y$-$x$ plane), and (b) view 2 ($45^{\circ}$ counter clockwise in the $y$-$x$ plane at $t =2.0$ ns, for $E_{p} = 5,~15,~25$ MeV protons.}
    \label{fig:radiographs}
\end{figure}

Synthetic radiographs of the FLASH-simulated plasmas have been produced by simulating $5$, $15$, and $25$ MeV protons with a beam divergence angle of $30^{\circ}$ from an infinitesimal source, using $5\times10^{6}$-$10^{7}$ protons in the beam, from two viewing angles: view 1 at $60^{\circ}$ clockwise from the $y$-axis, and view 2 at $45^{\circ}$ counter clockwise from the $y$-axis. These can be seen in figure \ref{fig:radiographs} (top and bottom row, respectively). 
For both viewing angles, the source and detector were placed $6$ mm and $130$ mm from the origin, respectively. TNSA proton yield is typically on the order of $10^{11}$ to $10^{13}$ particles \citep{macchi2013ion}, far exceeding that used here. However, the resolution and signal-to-noise ratio obtained in these simulations was satisfactory for our purposes. After they are generated, protons are instantaneously incident on a grid of equal dimensions to the FLASH simulation domain. Each grid vertex contains magnetic field data from the FLASH outputs at the selected time step. Proton deflections at each grid point are calculated using the Lorentz force, and the protons are advanced to the next grid point using a Boris pusher algorithm \citep{boris1970relativistic}. The magnetic field is assumed static during proton traversal. This assumption is justified since the protons' traversal time $\tau_{\rm p}$ through the plasma is much shorter than the evolution timescale $\tau_{B}$ of the magnetic field within the plasma ($\tau_{p} \sim 10^{-11}$ s $\ll \tau_{B}\sim 10^{-9}$ s). The temporal step size was set between $\delta x\left(2E_{p}/m_{p}\right)^{-1/2} \geq\delta t\geq 0.1\delta x\left(2E_{p}/m_{p}\right)^{-1/2}$, where $E_{p}$ is the proton energy and $m_{p}$ is the proton mass, and $\delta x$ is the grid spacing. Once all protons have crossed the grid, they are all instantaneously advanced onto a virtual detector. The positions and orientations of the source and detector relative to the grid are defined such that a vector is drawn from the origin of the grid to the source and the detector plane, determining beam direction. The target's shadows from the drive and shock foils were introduced into the radiographs by identifying the regions of solid density $n_{e} \gtrsim 10^{23}$ cm from the FLASH simulation data, and removing all protons that passed through such regions from the radiograph. The top rectangular region with no fluence evident in figure \ref{fig:radiographs} is the drive foil's shadow, and the lower is the shock foil's shadow. The small circle below the shock foil is a fiducial added for orientation purposes.

The proton radiographs themselves reveal qualitative features of the magnetic fields. The filamentary structures observed in figure \ref{fig:radiographs} have a much greater proton fluence than the mean, while voids have a much lower fluence. Given the specifics of proton-radiography set-up, this is consistent with the presence of multi-kilogauss magnetic fields in the plasma~\citep{bott2017proton}. The orientations of the filaments and voids directly above the shock foil's shadow are preferentially aligned with its upper surface, implying that the magnetic fields in the plasma are too. Another key signature of magnetic fields in the images is the up-down asymmetry in the global morphology of the proton radiographs for the two views. From view 1, void regions are seen directly above the shock foil, with darker filaments above the voids, implying that protons travelling into the shocked plasma have been deflected upwards, perpendicular to the direction of the magnetic field. View 2 shows the opposite morphology: the dark filaments and voids have swapped positions with their counterparts. This effect is because the direction of the magnetic field is reversed in the reference frame of the protons in this view, with the consequence that they have been deflected downwards instead. 

\begin{figure}
    \centering
    \includegraphics[width=1\linewidth]{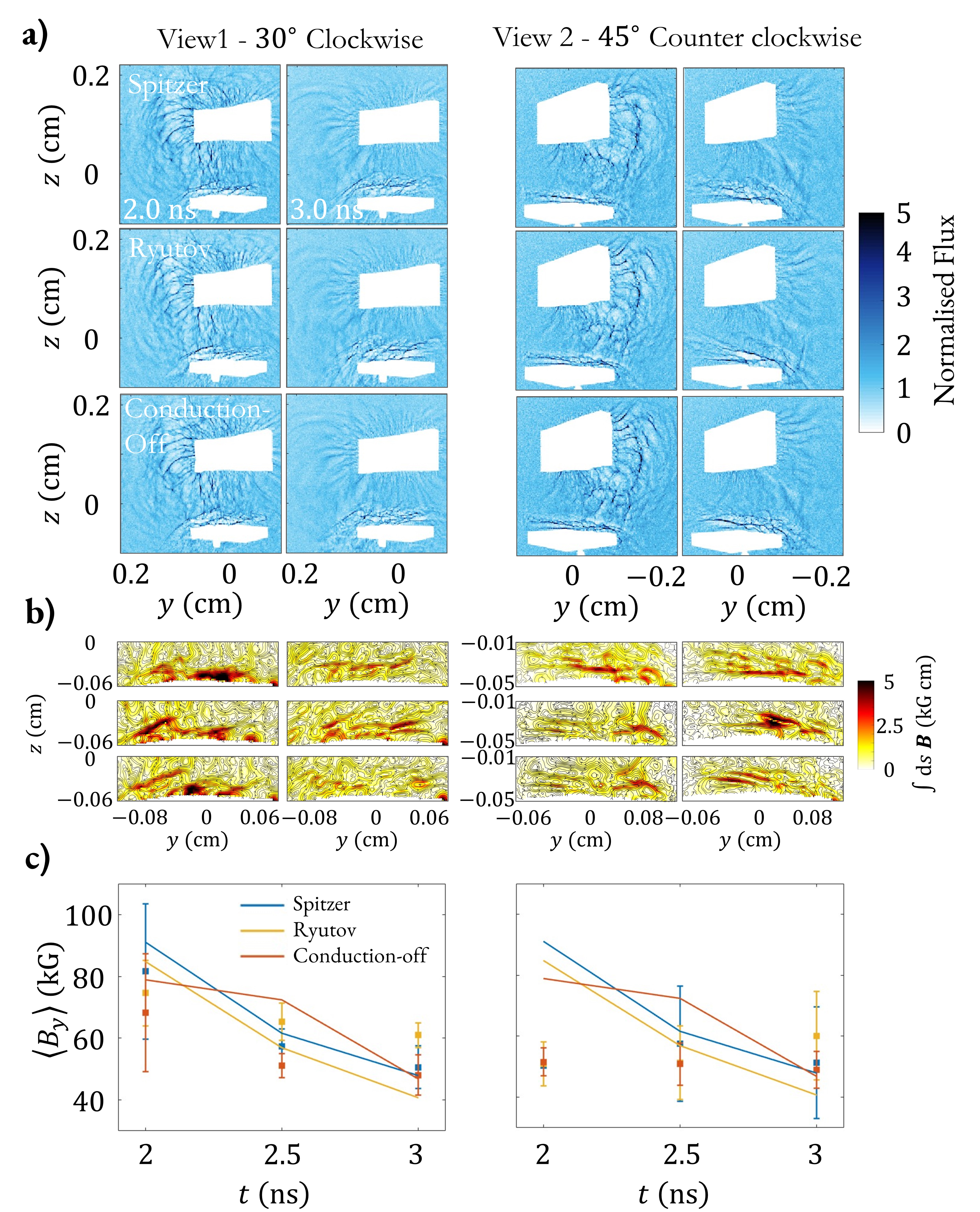}
    \caption{(a) Normalised 15-MeV proton fluence for each conduction model evolving in time from right to left between $2.0~\text{ns}\leq t\leq 3.0~\text{ns}$ for each detector view. (b) Reconstructed path integrated magnetic field directly above the shock foil, for each conduction model, similarly evolving in time as in (a), with arrows indicating mean direction of the field. (c) Time evolution of inferred magnetic field strength for each conduction model, and each detector view (left to right as in (a) and (b)), with the blue, yellow, and orange lines corresponding to the Spitzer, Ryutov, and conduction-off simulations, respectively.}
    \label{fig:prad_AllCondModels}
\end{figure}

More quantitatively, reconstructions of the path-integrated magnetic field $\int \mathrm{d}s \, \boldsymbol{B}$ can be obtained by comparing the detected proton fluence profile to a reference fluence profile obtained assuming no magnetic fields are present. Such reconstructions are well-defined mathematical inversion problems provided that proton trajectories do not self-intersect prior to reaching the detector due to deflections, giving rise to caustic structures. In the $5$ MeV radiographs, branched caustic structures~\citep{kugland2012invited} in the filaments directly above the shock foil shadow are clearly evident, implying that  magnetic field strengths reconstructed from these radiographs could be smaller than the true values. However, in the $25~\text{MeV}$ images, these caustic features are barely formed, implying that inversion should be possible. In the absence of caustics, the relation between the unperturbed and perturbed fluence profiles can be described by a Monge-Ampere equation, which can be efficiently inverted~\citep{dean2006numerical, sulman2011efficient}. For the purpose of carrying out the reconstruction, fluence images were generated on the simulation grid in the absence of any magnetic fields while retaining the target shadow for both viewing angles.
A small region was selected for the reconstruction -- the same region selected for both no-fluence and fluence profiles -- to focus on the fluence profile directly above the shock foil. The PROBLEM solver was then applied to produce the $\int \mathrm{d}s \, \boldsymbol{B}$ reconstructions shown in figure \ref{fig:prad_AllCondModels}~\citep{bott2017proton}. 

A time series of the evolution of simulated 15-MeV proton radiographs from both views, along with path-integrated magnetic-field reconstructions, are shown in figure \ref{fig:prad_AllCondModels} for the FLASH simulations.  
The fluence images appear qualitatively similar, irrespective of the conduction model adopted: all exhibit voids and filaments above the shock foil, and as time progresses, their spatial extent enlarged, though greater stochasticity is seen in the Ryutov, and conduction-off simulations. The reconstructed path-integrated magnetic fields are also similar, but not exactly so. Their magnitudes are comparable, but their overall morphology begins to deviate as time progresses. The simulations with Spitzer or Ryutov conduction models yield similar results throughout in strength and preferred direction. By $3.0~\text{ns}$ in the simulations with no conduction, the magnetic field has completely broken up, deviating from the results of the other conduction models. The similarity of the reconstructed magnetic fields between the Ryutov and Spitzer models is perhaps surprising, given the much larger magnetic field perturbations present in the FLASH simulations with Ryutov conduction (see figure \ref{fig:magFieldMorphology}(b)). The minimal effect of these perturbations on the trajectory of the protons is plausibly explained by many of these perturbed field lines being approximately parallel to the flight path of the protons, significantly reducing deflections due to the Lorentz force. These results suggest that any differences in how the magnetic field may present itself, based on the properties of thermal conduction in the plasma, will be more distinct at later times in the plasma's evolution.

To infer characteristic magnetic field strengths themselves, as opposed to path-integrated values, we assume that the magnetic field lines do not reverse sign along the path of the protons, and therefore, $\langle \boldsymbol{B} \rangle_{\text{LoS}} \approx \left(\int \mathrm{d}s \, \boldsymbol{B}\right)/\ell_{\rm path}$, 
where $\ell_{\rm path}$ is the path length of the protons through the plasma. $\ell_{\rm path}$ is, in turn, related to 
the (known) length $\ell_{x}$ of the shock foil along $x$, and the angle $\alpha$ from the LoS of the proton beam to the $y$-axis, via $\ell_{\rm path} \approx \ell_{x}/\sin{\alpha}$.
Since the $y$-component of the field, $\langle {B}_{y} \rangle_{\text{LoS}}$, is much larger than the other components, at least in the Spitzer simulations, and the protons travel at an angle $\alpha$ with respect to this component, the effective deflection field experienced by the protons is $\langle {B}_{y} \rangle_{\text{LoS}} \sin{\alpha}$. Therefore, the estimated field strength (along $y$) can be inferred from the reconstructed path-integrated magnetic fields by $\langle {B}_{y} \rangle_{\text{LoS}} \approx (\int \mathrm{d}s \, \boldsymbol{B})/\ell_{x}$.

We test the efficacy of such an inference by comparing $\langle {B}_{y} \rangle_{\text{LoS}}$ computed in this manner to the average values of $B_{y}$ computed directly from the FLASH simulations. The field values inferred from the reconstructed, path-integrated magnetic-field maps were obtained by averaging over a $0.2~\text{cm}\times0.02~\text{cm}$ box (approximate size of shocked plasma), accounting for the foreshortening effects due to the respective viewing angles of each image. These were then compared to the FLASH values averaged, post Mach number truncation. The centre position of the box was adjusted between each time and view in order to capture the field strength that overlaps with the position of the shocked plasma. The path length through the plasma was $\ell_{x} =0.02~\text{cm}$. The errors were calculated by splitting the reconstructed field data into three equally sized boxes, covering the extent of the field of interest, and taking the standard deviation between the average within each box. 

The results of this comparison, shown in figure \ref{fig:prad_AllCondModels}(c), are in good agreement for view $1$, particularly for the Spitzer simulations. View $2$ shows considerable underpredictions at $2.0~\text{ns}$, but converge with the FLASH results as time passes. This deviation is primarily due to the morphology of the fluence profiles during the reconstruction of the fields. As discussed, the dark filamentary fluence structures for view $2$ are located directly above the shock foil shadow at early times, making it difficult to isolate the full extent of the fluence structure, leading to erroneous field reconstructions. This is not a problem at later times as the expanding plasma will advect the magnetic field, therefore the fluence filaments, upwards. In the conduction-off simulations another contributor is likely the stochasticity of the field, which can be known to produce lower field values for this LoS integrated diagnostic \citep{bott2017proton}. Underpredictions are also commonly attributed to caustic structures; to test this hypothesis, proton radiographs of the Spitzer-FLASH simulations at $2.0~\text{ns}$ for view $2$ have been generated using $35~\text{MeV}$ and $50~\text{MeV}$ protons (not shown) as well as $15~\text{MeV}$. The resulting field estimates showed minimal deviation from the plotted values. The same test was performed for the conduction-off simulations at $2.5~\text{ns}$ also using $35~\text{MeV}$ and $50~\text{MeV}$ protons (not shown), with results also not showing significant deviation from the plotted values. This shows that despite some caustic broadening being observed in the fluence profiles, the effect on the reconstructed magnetic fields is negligible when compared to the other contributor of the underpredictions. Finally, in this analysis, the spatial region was selected based on notable structure location within the reconstructed fields. If this region is not carefully selected it could result in significant contributions from the jet reducing the overall average. Observations from the x-ray self-emission will be a good indicator for region selection, and expected structure size during the analysis of the experimental data. 

\section{Discussion \& Conclusions}\label{sec:discussion}
Through multiple simulation campaigns using the MHD code FLASH, we have demonstrated the feasibility of an experimental platform capable of generating a planar plasma with sustained temperature gradients, threaded with magnetic fields aligned parallel to those gradients. We have presented a set of primary diagnostics intended for use during the upcoming Orion campaign, supported by synthetic outputs to aid in our understanding of the characteristic signals we expect to observe and have informed decisions on diagnostic timing and placement. With the diagnostics presented, we have shown we can infer key plasma properties, namely the electron temperature and gradient scales, and the magnetic field strength and direction. Successful examples of such inference have been provided via comparisons between post-processed synthetic diagnostic outputs and values taken directly from the FLASH simulation.

\begin{table}
  \begin{center}
\def~{\hphantom{0}}
  \begin{tabular}{lcccc}
\multicolumn{1}{l}{\textbf{\underline{Spitzer}}} \\

 Quantity & $1.5$ ns & $2.0$ ns & $2.5$ ns & $3.0$ ns \\
 \hline

 $T_{e}$ (eV) & $1440$ & $606$ & $353$ & $226$ \\

 $n_{e}$ ($10^{18}$ cm$^{-3}$) & $3.34$ & $8.05$ & $8.44$ & $7.59$ \\

 $B$ (kG) & 119 & 117 & 86 & 60 \\
 
 $L_{T}$ (cm) & $ 0.21 $ & $0.51$ & $0.69$ & $3.39$ \\ 

 $\lambda_{e}$ (cm) & $0.30$ & $0.027$ & $0.010$ & $ 0.006 $ \\

 $\rho_{e}$ (cm) & $1.4\times10^{-3}$ & $0.8\times10^{-3}$ & $0.8\times10^{-3}$ & $0.9\times10^{-3}$ \\
 
  $\text{Ha}_{e}^{-1}$ & $0.005$ & $0.029$ & $0.078$ & $0.16$ \\

  $\beta_{e}$ & $56$ & $39$ & $35$ & $41$ \\

  $\tau_{\chi_{e}}$ (s) & $ 7.4\times10^{-12}$ & $8.0\times10^{-10}$ & $5.4\times10^{-9}$ & $2.9\times10^{-7}$ \\
 
  $q_{||\text{eff}}/q_{S}$ & $ 0.20 $ & $0.63$ & $0.86$ & $0.98$ \\
\hline
  \multicolumn{1}{l}{ } \\
\multicolumn{1}{l}{\textbf{\underline{Ryutov}}} \\

 Quantity & $1.5$ ns & $2.0$ ns & $2.5$ ns & $3.0$ ns \\
 \hline
 
 $T_{e}$ (eV) & $3380$ & $2000$ & $1390$ & $936$ \\

 $n_{e}$ ($10^{18}$ cm$^{-3}$) & $2.73$ & $7.83$ & $7.43$ & $5.37$ \\

 $B$ (kG) & 123 & 132 & 113 & 83 \\
 
 $L_{T}$ (cm) &$ 0.052 $ & $ 0.066 $ & $ 0.22 $ & $ 0.89 $ \\ 

 $\lambda_{e}$ (cm) &$ 1.71 $ & $ 0.25 $ & $ 0.10 $ & $ 0.07 $ \\

 $\rho_{e}$ (cm) & $ 2.2\times10^{-3}$ & $ 2.0\times10^{-3}$ & $ 1.4\times10^{-3}$ & $ 1.6 \times10^{-3}$ \\
 
  $\text{Ha}_{e}^{-1}$ & $ 0.001$ & $ 0.008$ & $ 0.014$ & $ 0.023 $ \\

  $\beta_{e}$ & $118$ & $192$ & $90$ & $71$ \\

  $\tau_{\chi_{e}}$ (s) & $4.0 \times10^{-11}$ & $ 9.1 \times10^{-11} $ & $ 1.8 \times10^{-9}$ & $ 3.4 \times10^{-8}$ \\
 
  $q_{||\text{eff}}/q_{S}$ & $ 0.093 $ & $ 0.063 $ & $ 0.17 $ & $ 0.42 $ \\
  \hline
  \multicolumn{1}{l}{ } \\
\multicolumn{1}{l}{\textbf{\underline{Conduction-off}} $^{\dagger}$} \\

 Quantity & $1.5$ ns & $2.0$ ns & $2.5$ ns & $3.0$ ns \\
 \hline
 
 $T_{e}$ (eV) & $6350$ & $3610$ & $1780$ & $1070$ \\

 $n_{e}$ ($10^{18}$ cm$^{-3}$) & $1.96$ & $5.69$ & $6.00$ & $5.13$ \\

 $B$ (kG) & 125 & 143 & 102 & 73 \\
 
 $L_{T}$ (cm) &$ 0.061 $ & $ 0.15 $ & $ 0.12 $ & $ 0.25 $ \\ 

 $\lambda_{e}$ (cm) &$ 8.97 $ & $ 1.2 $ & $ 0.24 $ & $ 0.11 $ \\

 $\rho_{e}$ (cm) & $ 2.2\times10^{-3}$ & $ 1.4\times10^{-3}$ & $ 1.2\times10^{-3}$ & $ 1.3 \times10^{-3}$ \\
 
  $\text{Ha}_{e}^{-1}$ & $ 0.2\times10^{-3}$ & $ 0.001$ & $ 0.005$ & $ 0.011 $ \\

  $\beta_{e}$ & $92$ & $69$ & $63$ & $59$ \\
  
  $\tau_{\chi_{e}}$ (s) & - & - & - & - \\
  
 
  $q_{||\text{eff}}/q_{S}$ & $ 0.12 $ & $ 0.15 $ & $ 0.18 $ & $ 0.24 $ \\
  \hline
  \end{tabular}
    \caption{Plasma parameters of interest calculated from the FLASH simulation data with Spitzer (top), Ryutov (centre), and conduction-off (bottom). These values were obtained using the coefficients and formulae from table \ref{table:equations_for_calcs}. The Mach number truncation averaging procedure was used on each parameter, and then averaged over the to obtain these values, see section \ref{sec:sim_plasma_param}. For $L_{T}$ a smoothing procedure, similar to the one described for figure \ref{fig:magFieldMorphology}, was used taking into account regions with monotonic gradients.\\
    \newline
    $^{\dagger}$ For the conduction-off simulations the heat-flux, and thermal-diffusive terms are to be considered in the case of a Spitzer governed plasma which spontaneously becomes non conductive at the reported time-step, and are given to illustrate a maximum boundary on the predicted parallel suppression.}
    \label{table:plasma_parameters}
  \end{center}
\end{table}
 \begin{figure}
    \centering
    \includegraphics[width=1\linewidth]{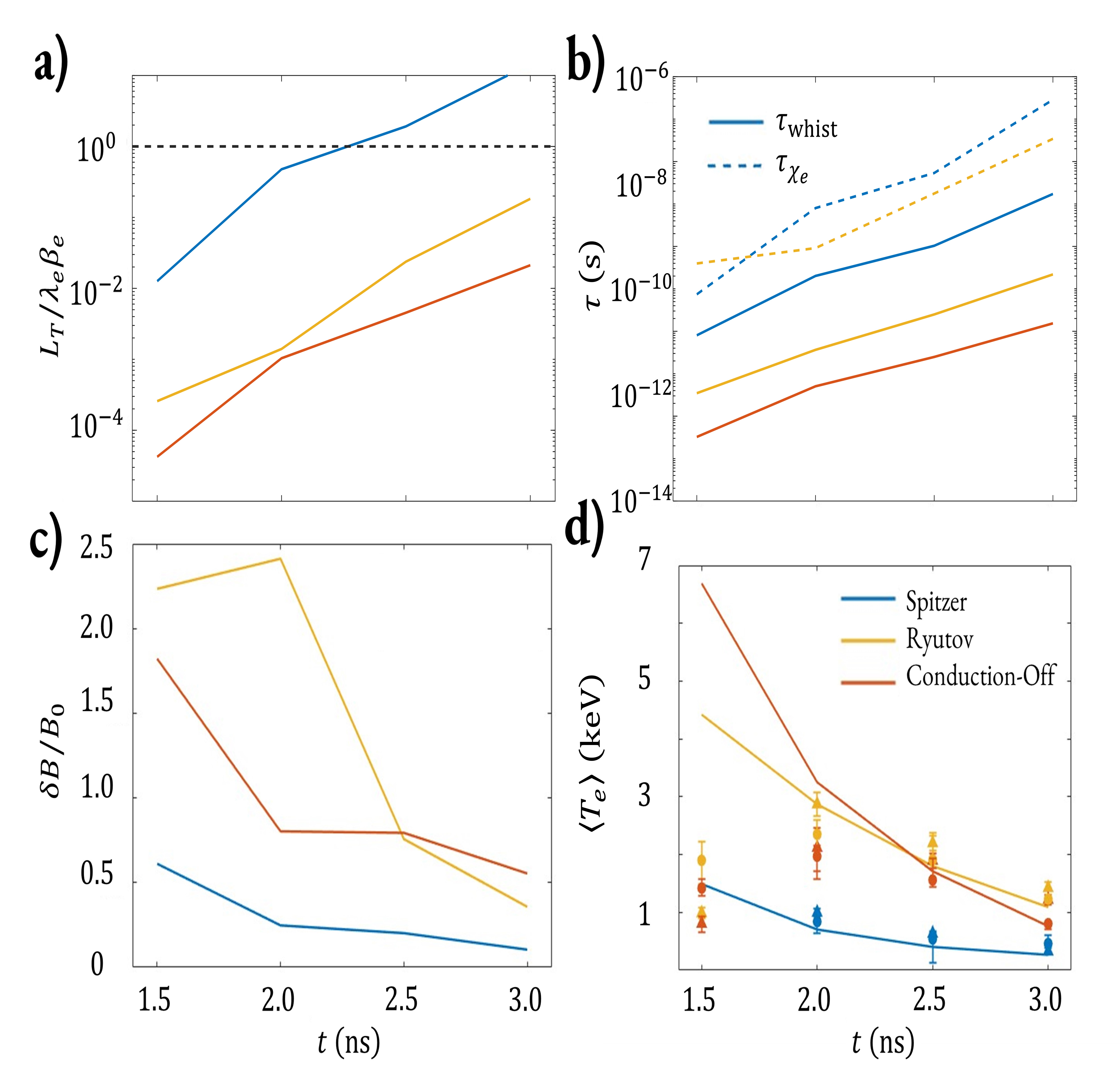}
    \caption{In the above figure the blue, yellow, and orange lines correspond to results from simulations with Spitzer, Ryutov, and conduction-off conductivity, respectively. (a) Evolution of the instability condition, i.e. the whistler heat-flux instability can grow when below unity (dashed line). (b) Saturation timescale of the instability (solid lines) compared to the thermal diffusive timescale of the plasma at the same time (dashed line). Note the absence of the diffusive timescale for the conduction-off simulations, this is because there is no conduction, and therefore will have no thermal diffusivity. (c) Relative amplitude of the whistler inferred from the $(\delta B/B_{0})^{2}\sim\beta_{e}\rho_{e}/L_{T}$ relation. It should be noted that all of these calculations have been done using macroscopic quantities averaged from FLASH. (d) Temperature evolution between $1.5~\text{ns}\leq t\leq\text{ns}$ of the shocked plasma for each conduction model (solid lines), overlaid with the synthetic x-ray diagnostic inferred $\langle T_{e}\rangle$ using post processed FLASH simulation data. The triangles correspond to spectroscopy inferred values, and the circles from the GXD ratio curve analysis.}
    \label{fig:WhistlerScaleComp}
\end{figure}
 One key question to consider is whether the whistler heat-flux instability is, indeed, expected to act in the plasma produced in our planned experiment. Using $T_{e}$, $n_{e}$, and $|\mathbf{B}|$ taken directly from FLASH, we can compute several key theoretical quantities to assess whether the plasma is susceptible to the whistler-heat flux instability. The results are summarised in table \ref{table:plasma_parameters}, with relevant coefficients and the equations used for these calculations seen in table \ref{table:equations_for_calcs}. For the whistler heat-flux instability to have a noticeable effect on the plasma's thermal conduction, several conditions must be satisfied (see section \ref{sec:design_overview}). As shown in Figure \ref{fig:magFieldMorphology}, there exist temperature gradients parallel to magnetic field lines throughout the plasma, a necessary condition of instability. The condition for the plasma to become unstable to the whistler heat-flux instability, $L_{T}/\lambda_{e}\lesssim \beta_{e}$, also needs to be satisfied~\citep{bott2024kinetic}. Figure \ref{fig:WhistlerScaleComp}a shows a plot of $L_{T}/\lambda_{e}\beta_{e}$ between $1.5~\text{ns}\leq t \leq 3.0~\text{ns}$. It can be seen that, with the exception of the Spitzer model, this condition is satisfied across the whole time domain with Spitzer only being susceptible for $t \leq 2.5~\text{ns}$. The instability must also have sufficient time to grow; to assess this, we compare the linear-growth timescale of the instability, $\tau_{\text{whist}}\approx \big\{ 0.56\lambda_{e}/L_{T}\big[ 1 - 0.13(\beta_{e}\lambda_{e}/L_{T})^{-2/5} \big] \Omega_{e} \big\}^{-1}$ \citep{bott2024kinetic}, with the Spitzer electron thermal diffusive timescale, $\tau_{\chi_{e}}\sim L_{T}^{2}/\chi_{e}$. These are shown in figure \ref{fig:WhistlerScaleComp}b, indicated by the solid and dashed lines respectively. Across the whole time domain the linear-growth timescale is shorter than the diffusive timescale for Spitzer conduction by at least an order of magnitude, and by two orders of magnitude for Ryutov conduction, with the saturation timescale for the conduction-off simulations sharing a similar evolution with Ryutov. This demonstrates the instability can grow before the plasma can reach thermal equilibrium, and that this growth is substantially faster than the rate at which the plasma evolves dynamically. 

 Finally, for all conduction models across the entire time domain, the electron gyroradius is much smaller that the macroscopic temperature gradient length: $\rho_{e}\ll L_{T}$. Because the characteristic length scale of the instability is $\mathcal{O}(\rho_{e})$, it follows that the whistler heat-flux instability can grow in the generated plasma, regardless of the conduction model, provided the required alignment between temperature gradient to magnetic field is present.
 This holds in spite of the collisionality condition, $\lambda_{e}<L_{T}$, not being met at early times in the Ryutov and conduction-off cases: the whistler heat-flux instability can be driven by free-streaming particles in quasi-collisionless plasmas. 
 
Having established that the whistler heat-flux instability should operate, we examine its expected effect on heat conduction. As discussed in the introduction, this is related to the expected amplitude of the magnetic fluctuations associated with the whistler heat-flux instability. These are estimated using a relation $(\delta B/B_{0})^{2}\sim\beta_{e}\rho_{e}/L_{T}$ derived from previous theory and computational studies~\citep{komarov2018self, yerger2025collisionless}. The evolution of the predicted whistler amplitudes is shown in figure \ref{fig:WhistlerScaleComp}c, across the same time domain as in figure \ref{fig:WhistlerScaleComp}a and b, revealing significant magnetic perturbations at early times. This supports the relevance of suppressed conduction models such as the Ryutov model. These perturbations act as indicators of the isotropisation of the heat flux along magnetic field lines. As expected, decreasing fluctuations over time correspond with the increase of the suppression factor $q_{||\text{eff}}/q_{\text{S}}$, as shown in table \ref{table:plasma_parameters}. 

\begin{table}
  \begin{center}
  \renewcommand{\arraystretch}{1.5}
\def~{\hphantom{0}}
  \begin{tabular}{lccc}

         Quantity & Formula \\
         \hline
         
        Sound speed ($c_{\text{s}}$) & $\sqrt{ \gamma\left( Z_{\text{q,eff}}T_{e} + T_{i}\right)/1.67\times10^{-24}m_{\text{eff}}}$  \\
        
         Mach number ($M_{z}$) & $v_{z}/c_{\textbf{s}}$  \\
         
        Coulomb logarithm ($\log\Lambda$) & $ 23.5 - \log n_{e}^{1/2}T_{e}^{-5/4} - \sqrt{1\times10^{-5} + (\log(T_{e}) - 2)^{2}/16} $  \\
         
         Electron mean-free path ($\lambda_{e}$) & $2.1\times10^{13}T_{e}^{2}/Z_{\text{eff}}n_{e}\log\Lambda$ \\
         
        Electron gyroradius ($\rho_{e}$) & $2.4T_{e}^{1/2}|\boldsymbol{B}|^{-1}$ \\
        
        Temperature gradient length scale ($L_{T}$) & $\nabla_{||}\log T_{e}$  \\
        
        Electron plasma beta ($\beta_{e}$) & $4\times10^{-11}n_{e}T_{e}|\boldsymbol{B}|^{-2}$  \\
        
        Inverse Hall parameter ($\text{Ha}_{e}^{-1}$) & $\rho_{e}/\lambda_{e}$  \\
        
        Thermal diffusivity ($\chi_{e}$) & $3.1\times10^{21}T_{e}^{5/2}/ Z_{\text{eff}}n_{e}\log\Lambda $  \\
         
         Thermal diffusive timescale ($\tau_{\chi_{e}}$) & $L_{T}^{2}/\chi_{e}$ (Spitzer), $\text{Ha}_{e}L_{T}^{2}/\chi_{e}$ (Ryutov)\\
         
         Parallel heat flux suppression ($q_{||\text{eff}}/q_{||\text{S}}$) & $\left[ 1+\beta_{e}/3\left( L_{T}/\lambda_{e} + 4\right)\right]^{-1}$  \\
  \end{tabular}
  \caption{Theoretical plasma parameters used for analysis throughout, formulae in CGS units with temperature in eV. Numerical coefficients used above are the following, for a parylene-C plasma: effective ion charge $Z_{\text{eff}} = 7.7$, $Z_{\text{q,eff}} = 4.44$ using quasineutrality, average atomic weight $m_{\text{eff}} = 7.80$, and adiabatic index $\gamma = 5/3$. The parameter expressions can be found in \cite{bott2021time}.}
  \label{table:equations_for_calcs}
  \end{center}
\end{table}

We have established that the plasma is not only susceptible to the whistler heat-flux instability, but the relevant timescales indicate the instability can fully saturate before significant thermal evolution occurs. The platform has been designed to generate self-magnetised plasmas, and replicating similar plasma conditions in the absence of magnetic fields would be challenging without imposing potentially perturbative external fields. This leads to difficulties when considering the thermal evolution of such a plasma with no fields present without the use of simulations. Potential alterations to the platform that would enable such studies will be explored for future campaigns.

In summary, early in its evolution the plasma attains a state which is susceptible to the whistler heat-flux instability, with a predicted maximum suppression of the effective heat-flux $(q_{||\text{eff}}/q_{\text{S}})^{-1}\approx16.4$ relative to the Spitzer value. As the plasma cools, this suppression factor approaches unity, indicating that at lower temperatures the conduction physics becomes more Spitzer-like. Therefore, it becomes increasingly difficult to distinguish between a Spitzer governed plasma with that governed by a suppressed conduction model. That being said, it has been shown through a series of synthetic diagnostic analyses that there are significant, measurable, characteristics in the evolution of the plasma that can give indications to the conduction model present, as illustrated in figure \ref{fig:WhistlerScaleComp}. This is particularly evident between $2.0\leq t\leq 2.5~\text{ns}$, where the inferred $T_{e}$ of the plasma from the synthetic diagnostics for the alternative conduction models are clearly distinct from those of the Spitzer plasma (figure \ref{fig:WhistlerScaleComp}d). Future simulation work will involve additional simulation campaigns introducing anisotropic heat flux governed by Braginskii transport, as well as models more analogous to suppressed conduction described by equation \ref{eq:komarovHeatFlux}. These developments will be invaluable when comparing the simulation outputs with the upcoming experimental data.

\section*{Acknowledgements}
The research leading to these results has received funding from AWE plc. A.F.A.B. was supported for this research by funding from the UKRI (grant number MR/W006723/1). The work of G.G. was supported in part by the UK Research and Innovation (UKRI) Frontier Research Guarantee under Grant No EP/Y035038/1 and by Science and Technology Facilities Council (STFC) under Grant No ST/W000903/1. We acknowledge support by the U.S. DOE NNSA under Awards DE-NA0002724, DE-NA0003605, DE-NA0003934, DE-NA0004144, DE-NA0004147, and Subcontracts 630138 and C4574 with LANL; the U.S. DOE Office of Science under Award DE-SC0021990; and the NSF under Awards PHY-2033925 and PHY-2308844. The software used in this work was developed in part by the U.S. DOE NNSA- and U.S. DOE Office of Science-supported Flash Center for Computational Science at the University of Chicago and the University of Rochester. UK Ministry of Defence © Crown owned copyright 2026/AWE. The authors would like to thank Henry Edwards at the Central Laser Facility for their contribution to target fabrication.

Competing interests: The authors declare none.

\bibliographystyle{jpp}

\bibliography{references}

@Article{abu2022lawson,
  author = {Abu-Shawareb, H., et al.},
  title = {Lawson criterion for ignition exceeded in an inertial fusion experiment},
  journal = {Phys. Rev. Lett.},
  year = {2022},
  volume = {129},
  pages = {075001},
}

@Article{spitzer1953transport,
  author = {Spitzer Jr, L. and H{\"a}rm, R.},
  title = {Transport phenomena in a completely ionized gas},
  journal = {Phys. Rev.},
  year = {1953},
  volume = {89},
}

@Article{cohen1950electrical,
  author = {Cohen, R. S. and Spitzer Jr, L. and Routly, P. M.},
  title = {The electrical conductivity of an ionized gas},
  journal = {Phys. Rev.},
  year = {1950},
  volume = {80},
  pages = {230},
}

@Article{komarov2018self,
  author = {Komarov, S. V. and Schekochihin, A. A. and Churazov, E. and Spitkovsky, A.},
  title = {Self-inhibiting thermal conduction in a high-{$\beta$}, whistler-unstable plasma},
  journal = {J. Plasma Phys.},
  year = {2018},
  volume = {84},
}

@Article{bott2024kinetic,
  author = {Bott, A. F. A. and Cowley, S. C. and Schekochihin, A. A.},
  title = {Kinetic stability of {C}hapman-{E}nskog plasmas},
  journal = {J. Plasma Phys.},
  year = {2024},
  volume = {90},
  pages = {975900207}
}

@Article{ryutov1999similarity,
  author = {Ryutov, D. and Drake, R. P. and Kane, J. and Liang, E. and Remington, B. A. and Wood-Vasey, W. M.},
  title = {Similarity criteria for the laboratory simulation of supernova hydrodynamics},
  journal = {Astrophys. J.},
  year = {1999},
  volume = {518},
}

@article{danson2019petawatt,
  author={Danson, C. N., et al.},
  title={Petawatt and exawatt class lasers worldwide},
  journal={High Power Laser Sci. Eng.},
  year={2019},
  volume={7},
  pages = {e54},
}

@article{meinecke2022strong,
  author={Meinecke, J., et al.},
  title={Strong suppression of heat conduction in a laboratory replica of galaxy-cluster turbulent plasmas},
  journal={Sci. Adv.},
  year={2022},
  volume={8},
}

@article{bott2021time,
  author={Bott, A. F. A., et al.},
  title={Time-resolved turbulent dynamo in a laser plasma},
  journal={Proc. Natl. Acad. Sci. U.S.A},
  year={2021},
  volume={118},
  pages = {e2015729118},
}

@article{braginskii1965transport,
  author={Braginskii, S. I.},
  title={Transport processes in a plasma},
  journal={Rev. Plasma Phys.},
  year={1965},
  volume={1},
 pages = {205},
}

@article{chang2011fusion,
  author={Chang, P. Y. and Fiksel, G. and Hohenberger, M. and Knauer, J. P. and Betti, R. and Marshall, F. J. and Meyerhofer, D. D. and S{\'e}guin, F. H. and Petrasso, R. D.},
  title={Fusion yield enhancement in magnetized laser-driven implosions},
  journal={Phys. Rev. Lett.},
  year={2011},
  volume={107},
  pages = {035006},
}

@article{walsh2017self,
  author={Walsh, C. A. and Chittenden, J. P. and McGlinchey, K. and Niasse, N. P. L. and Appelbe, B. D.},
  title={Self-generated magnetic fields in the stagnation phase of indirect-drive implosions on the {N}ational {I}gnition {F}acility},
  journal={Phys. Rev. Lett.},
  year={2017},
  volume={118},
}

@article{moody2022increased,
  author={Moody, J. D., et al.},
  title={Increased ion temperature and neutron yield observed in magnetized indirectly driven d 2-filled capsule implosions on the national ignition facility},
  journal={Phys. Rev. Lett.},
  year={2022},
  volume={129},
}

@article{fabian1994cooling,
  author={Fabian, A. C.},
  title={Cooling flows in clusters of galaxies},
  journal={Annu. Rev. Astron. Astrophys},
  year={1994},
  volume={32},
}

@article{komarov2016thermal,
  author={Komarov, S. V. and Churazov, E. M. and Kunz, M. W. and Schekochihin, A. A.},
  title={Thermal conduction in a mirror-unstable plasma},
  journal={Mon. Not. R. Astron. Soc.},
  year={2016},
  volume={460},
}

@article{henchen2018observation,
  author={Henchen, R J and Sherlock, M and Rozmus, W and Katz, J and Cao, D and Palastro, J P and Froula, D H},
  title={Observation of nonlocal heat flux using {T}homson scattering},
  journal={Phys. Rev. Lett.},
  year={2018},
  volume={121},
}

@article{komarov2014suppression,
  author={Komarov, S. V. and Churazov, E. M. and Schekochihin, A. A. and ZuHone, J. A.},
  title={Suppression of local heat flux in a turbulent magnetized intracluster medium},
  journal={Mon. Not. R. Astron. Soc.},
  year={2014},
  volume={440},
}

@article{roberg2018suppression,
  author={Roberg-Clark, G. T. and Drake, J. F. and Reynolds, C. S. and Swisdak, M.},
  title={Suppression of electron thermal conduction by whistler turbulence in a sustained thermal gradient},
  journal={Phys. Rev. Lett.},
  year={2018},
  volume={120},
}

@article{tzeferacos2018laboratory,
  author={Tzeferacos, P., et al.},
  title={Laboratory evidence of dynamo amplification of magnetic fields in a turbulent plasma},
  journal={Nature Comm.},
  year={2018},
  volume={9},
}

@article{hopps2015comprehensive,
  title={Comprehensive description of the {O}rion laser facility},
  author={Hopps, N., et al.},
  journal={Plasma Phys. Control. Fusion},
  year={2015},
  volume={57},
}

@article{fryxell2000flash,
  title={{FLASH}: An adaptive mesh hydrodynamics code for modeling astrophysical thermonuclear flashes},
  author={Fryxell, B., et al.},
  journal={Astrophys. J. Suppl.},
  year={2000},
  volume={131},
}

@article{nagayama2000theory,
  author={Nagayama, K.},
  title={Theory of Shock Waves},
  journal={Handbook of {S}hock {W}aves, {T}hree {V}olume {S}et},
  year={2000},
  pages={315, 338},
  publisher={Elsevier}
}

@book{colvin2013extreme,
  author={Colvin, J. and Larsen, J.},
  title={Extreme physics: properties and behavior of matter at extreme conditions},
  year={2013},
  pages = {127, 132},
  chapter = {5},
  publisher={Cambridge University Press}
}

@article{tubman2021observations,
  author={Tubman, E. R., et al.},
  title={Observations of pressure anisotropy effects within semi-collisional magnetized plasma bubbles},
  journal={Nature Comm.},
  year={2021},
  volume={12},
}

@article{fox2011fast,
  author={Fox, W. and Bhattacharjee, A. and Germaschewski, K.},
  title={Fast magnetic reconnection in laser-produced plasma bubbles},
  journal={Phys. Rev. Lett.},
  year={2011},
  volume={106},
}

@article{nilson2006magnetic,
  author={Nilson, P. M., et al.},
  title={Magnetic reconnection and plasma dynamics in two-beam laser-solid interactions},
  journal={Phys. Rev. Lett.},
  year={2006},
  volume={97},
}

@book{rybicki1991radiative,
    author = {Rybicki, G. B. and Lightman, A. P.},
    title = {Radiative processes in astrophysics},
    chapter = {5},
    pages = {162},
    year ={1991},
    publisher = {John Wiley \& Sons}
}

@article{presnyakov1976x,
  author={Presnyakov, L. P.},
  title={X-ray spectroscopy of high-temperature plasma},
  journal={Sov. Phys. Usp.},
  year={1976},
  volume={19},
}

@article{gabriel1972dielectronic,
  author={Gabriel, A. H.},
  title={Dielectronic satellite spectra for highly-charged helium-like ion lines},
  journal={Mon. Not. R. Astron. Soc.},
  year={1972},
  volume={160},
}

@article{macchi2013ion,
  author={Macchi, A. and Borghesi, M. and Passoni, M.},
  title={Ion acceleration by superintense laser-plasma interaction},
  journal={Rev. Mod. Phys.},
  year={2013},
  volume={85},
}

@article{bott2017proton,
  author={Bott, A. F. A. and Graziani, C. and Tzeferacos, P. and White, T. G. and Lamb, D. Q. and Gregori, G. and Schekochihin, A. A.},
  title={Proton imaging of stochastic magnetic fields},
  journal={J. Plasma Phys.},
  year={2017},
  volume={83},
  pages = {905830614},
}

@article{schaeffer2023proton,
  author={Schaeffer, D. B., et al.},
  title={Proton imaging of high-energy-density laboratory plasmas},
  journal={Rev. Mod. Phys.},
  year={2023},
  volume={95},
}

@article{kugland2012invited,
  author={Kugland, N. L. and Ryutov, D. D. and Plechaty, C. and Ross, J. S. and Park, H. S.},
  title={Invited article: {R}elation between electric and magnetic field structures and their proton-beam images},
  journal={Rev. Sci. Instrum},
  year={2012},
  volume={83},
}

@article{lea1973thermal,
  title={Thermal-Bremsstrahlung interpretation of cluster {X}-ray sources},
  author={Lea, S. M. and Silk, J. and Kellogg, E. and Murray, S.},
  journal={Astrophys. J.},
  volume={184},
  pages={L105},
  year={1973}
}

@article{fabian1977subsonic,
  title={Subsonic accretion of cooling gas in clusters of galaxies},
  author={Fabian, A. C. and Nulsen, P. E. J.},
  journal={Mon. Not. R. Astron. Soc.},
  volume={180},
  pages={479--484},
  year={1977},
  publisher={Oxford University Press Oxford, UK}
}

@article{cowie1977radiative,
  title={Radiative regulation of gas flow within clusters of galaxies-A model for cluster {X}-ray sources},
  author={Cowie, L. L. and Binney, J.},
  journal={Astrophys. J.},
  volume={215},
  pages={723--732},
  year={1977}
}

@article{mcdonald2013growth,
  title={The growth of cool cores and evolution of cooling properties in a sample of 83 galaxy clusters at 0.3< z< 1.2 selected from the {SPT}-{SZ} survey},
  author={McDonald, M., et al.},
  journal={Astrophys. J.},
  volume={774},
  pages={23},
  year={2013},
  publisher={IOP Publishing}
}

@article{mcdonald2018revisiting,
  title={Revisiting the cooling flow problem in galaxies, groups, and clusters of galaxies},
  author={McDonald, M. and Gaspari, M. and McNamara, B. R. and Tremblay, G. R.},
  journal={Astrophys. J.},
  volume={858},
  pages={45},
  year={2018},
  publisher={IOP Publishing}
}

@article{fabian2012observational,
  title={Observational evidence of active galactic nuclei feedback},
  author={Fabian, A. C.},
  journal={Annu. Rev. Astron. Astrophys},
  volume={50},
  pages={455--489},
  year={2012},
  publisher={Annual Reviews}
}

@article{ruppin2023redshift,
  title={Redshift {E}volution of the {F}eedback--{C}ooling {E}quilibrium in the {C}ore of 48 {SPT} {G}alaxy {C}lusters: {A} {J}oint {C}handra--{SPT}--{ATCA} {A}nalysis},
  author={Ruppin, F.},
  journal={Astrophys. J.},
  volume={948},
  pages={49},
  year={2023},
  publisher={IOP Publishing}
}

@article{conroy2008thermal,
  title={Thermal balance in the intracluster medium: {I}s {AGN} feedback necessary?},
  author={Conroy, C. and Ostriker, J. P.},
  journal={Astrophys. J.},
  volume={681},
  pages={151},
  year={2008},
  publisher={IOP Publishing}
}

@article{zuhone2016cold,
  title={Cold fronts: probes of plasma astrophysics in galaxy clusters},
  author={ZuHone, J. A. and Roediger, E.},
  journal={J. Plasma Phys.},
  volume={82},
  pages={535820301},
  year={2016},
  publisher={Cambridge University Press}
}

@article{kempf2025non,
  title={Non-linear saturation and energy transport in global simulations of magneto-thermal turbulence in the stratified intracluster medium},
  author={Kempf, J. M. and Rincon, F.},
  journal={Astron. Astrophys.},
  volume={694},
  pages={A25},
  year={2025},
  publisher={EDP Sciences}
}

@article{kunz2011thermally,
  title={A thermally stable heating mechanism for the intracluster medium: turbulence, magnetic fields and plasma instabilities},
  author={Kunz, M. W. and Schekochihin, A. A. and Cowley, S. C. and Binney, J. J. and Sanders, J. S.},
  journal={Mon. Not. R. Astron. Soc.},
  volume={410},
  pages={2446},
  year={2011},
  publisher={The Royal Astronomical Society}
}

@article{o2025burn,
  title={Burn propagation in magnetized high-yield inertial fusion},
  author={O'Neill, S. T. and Appelbe, B. D. and Crilly, A. J. and Walsh, C. A. and Strozzi, D. J. and Moody, J. D. and Chittenden, J. P.},
  journal={Phys. Plasmas},
  volume={32},
  year={2025},
  publisher={AIP Publishing}
}

@article{Walsh_2025,
doi = {10.1088/1741-4326/adb7f0},
url = {https://dx.doi.org/10.1088/1741-4326/adb7f0},
year = {2025},
month = {feb},
publisher = {IOP Publishing},
volume = {65},
pages = {036040},
author = {Walsh, C. A., et al.},
title = {Magnetized {ICF} implosions: non-axial magnetic field topologies},
journal = {Nucl. Fusion},
}

@article{walsh2021biermann,
  title={Biermann battery magnetic fields in {ICF} capsules: {T}otal magnetic flux generation},
  author={Walsh, C. A. and Clark, D. S.},
  journal={Phys. Plasmas},
  volume={28},
  year={2021},
  publisher={AIP Publishing}
}

@article{rinderknecht2018kinetic,
  title={Kinetic physics in {ICF}: present understanding and future directions},
  author={Rinderknecht, H. G. and Amendt, P. A. and Wilks, S. C. and Collins, G.},
  journal={Plasma Phys. Control. Fusion},
  volume={60},
  pages={064001},
  year={2018},
  publisher={IOP Publishing}
}

@article{nicastro2008missing,
  title={Missing baryons and the warm-hot intergalactic medium},
  author={Nicastro, F. and Mathur, S. and Elvis, M.},
  journal={Science},
  volume={319},
  pages={55},
  year={2008},
  publisher={American Association for the Advancement of Science}
}

@article{macfarlane2007spect3d,
  title={{SPECT3D}--{A} multi-dimensional collisional-radiative code for generating diagnostic signatures based on hydrodynamics and {PIC} simulation output},
  author={MacFarlane, J. J. and Golovkin, I. E. and Wang, P. and Woodruff, P. R. and Pereyra, N. A.},
  journal={High Energy Density Phys.},
  volume={3},
  pages={181},
  year={2007},
  publisher={Elsevier}
}

@article{albertazzi2018experimental,
  author = {Albertazzi, B., et al},
  title = {Experimental platform for the investigation of magnetized-reverse-shock dynamics in the context of {POLAR}},
  journal = {High Power Laser Sci. Eng.},
  publisher = {Cambridge University Press},
  year = {2018},
  volume = {6},
  pages = {e43},
}

@article{Fatenejad2013172,
  author = {Fatenejad, M., et al.},
  title = {Modeling {HEDLA} magnetic field generation experiments on laser facilities},
  journal = {High Energy Density Phys.},
  year = {2013},
  volume = {9},
  pages = {172},
  url = {http://www.sciencedirect.com/science/article/pii/S1574181812001280},
  doi = {http://dx.doi.org/10.1016/j.hedp.2012.11.002}
}

@article{moczulski2024numerical,
  title={Numerical simulations of laser-driven experiments of ion acceleration in stochastic magnetic fields},
  author={Moczulski, K., et al.},
  journal={Phys. Plasmas},
  volume={31},
  year={2024},
  publisher={AIP Publishing}
}

@article{tzeferacos2015flash,
  title={{FLASH} {MHD} simulations of experiments that study shock-generated magnetic fields},
  author={Tzeferacos, P., et al.},
  journal={High Energy Density Phys.},
  volume={17},
  pages={24},
  year={2015},
  publisher={Elsevier}
}

@article{rigon2019rayleigh,
  title={Rayleigh-{T}aylor instability experiments on the {LULI2000} laser in scaled conditions for young supernova remnants},
  author={Rigon, G., et al.},
  journal={Phys. Rev. E},
  volume={100},
  pages={021201},
  year={2019},
  publisher={APS}
}

@article{stamper1971spontaneous,
  title={Spontaneous magnetic fields in laser-produced plasmas},
  author={Stamper, J. A. and Papadopoulos, K. and Sudan, R. N. and Dean, S. O. and McLean, E. A. and Dawson, J. M.},
  journal={Phys. Rev. Lett.},
  volume={26},
  pages={1012},
  year={1971},
  publisher={APS}
}

@article{stamper1975faraday,
  title={Faraday-rotation measurements of megagauss magnetic fields in laser-produced plasmas},
  author={Stamper, J. A. and Ripin, B. H.},
  journal={Phys. Rev. Lett.},
  volume={34},
  pages={138},
  year={1975},
  publisher={APS}
}

@article{biermann1950ursprung,
  title={{\"U}ber den {U}rsprung der {M}agnetfelder auf {S}ternen und im interstellaren {R}aum (miteinem {A}nhang von {A}. {S}chl{\"u}ter)},
  author={Biermann, L.},
  journal={Z. Naturforsch. A},
  volume={5},
  pages={65},
  year={1950}
}

@article{alfven1943existence,
  title={On the existence of electromagnetic-hydrodynamic waves},
  author={Alfv{\'e}n, H.},
  journal={Ark. Mat. Astron. Fys.},
  volume={29},
  pages={1},
  year={1943}
}

@article{yerger2025collisionless,
  title={Collisionless conduction in a high-beta plasma: a collision operator for whistler turbulence},
  author={Yerger, E. L. and Kunz, M. W. and Bott, A. F. A. and Spitkovsky, A.},
  journal={J. Plasma Phys.},
  volume={91},
  pages={E20},
  year={2025},
  publisher={Cambridge University Press}
}

@phdthesis{poole2024Development,
    title = {Development of advanced diagnostics for high energy density plasmas},
    author = {Poole, H.},
    year = {2024},
    school = {University of Oxford},
}

@inproceedings{boris1970relativistic,
  title={Relativistic plasma simulation-optimization of a hybrid code},
  author={Boris, J. P., et al.},
  booktitle={Proc. Fourth Conf. Num. Sim. Plasmas},
  pages={3},
  year={1970}
}

@article{bott2021inefficient,
  title={Inefficient magnetic-field amplification in supersonic laser-plasma turbulence},
  author={Bott, A. F. A., et al.},
  journal={Phys. Rev. Lett.},
  volume={127},
  pages={175002},
  year={2021},
  publisher={APS}
}

@article{sulman2011efficient,
  title={An efficient approach for the numerical solution of the {M}onge--{A}mp{\`e}re equation},
  author={Sulman, M. M. and Williams, J. F. and Russell, R. D.},
  journal={Appl. Numer. Maths},
  volume={61},
  pages={298},
  year={2011},
  publisher={Elsevier}
}

@article{dean2006numerical,
  title={Numerical methods for fully nonlinear elliptic equations of the {M}onge--{A}mp{\`e}re type},
  author={Dean, E. J. and Glowinski, R.},
  journal={Comput. Meth. Appl. Mech. Engng},
  volume={195},
  pages={1344},
  year={2006},
  publisher={Elsevier}
}

@article{valenzuela2024x,
  title={X-ray imaging and electron temperature evolution in laser-driven magnetic reconnection experiments at the national ignition facility},
  author={Valenzuela-Villaseca, V., et al.},
  journal={Phys. Plasmas},
  volume={31},
  year={2024},
  publisher={AIP Publishing}
}

@article{plasmapyCommunity_2024,
  author    = {{PlasmaPy Community}},
  title     = {PlasmaPy: Open-source software for plasma physics},
  volume   = {2024.10.0},
  year      = {2024},
  publisher = {Zenodo},
  journal       = {10.5281/zenodo.14010450}
}

@article{poole2026Git,
    author = {Poole, H.},
    title = {{XRFC} Ratiocurve, \url{https://github.com/Pooley12/XRFC_Ratiocurve.git}},
    year = {2026}
}

@ARTICLE{Pistinner_1998,
       author = {{Pistinner}, S.~L. and {Eichler}, D.},
        title = "{Self-inhibiting heat flux}",
      journal = {Mon. Not. R. Astron. Soc.},
         year = 1998,
        month = nov,
       volume = {301},
       number = {1},
        pages = {49-58},
          doi = {10.1046/j.1365-8711.1998.01770.x},
archivePrefix = {arXiv},
       eprint = {astro-ph/9807025},
 primaryClass = {astro-ph},
       adsurl = {https://ui.adsabs.harvard.edu/abs/1998MNRAS.301...49P}
}

@ARTICLE{Levinson_1992,
       author = {{Levinson}, Amir and {Eichler}, David},
        title = "{Inhibition of Electron Thermal Conduction by Electromagnetic Instabilities}",
      journal = {Astrophys. J.},
         year = 1992,
        month = mar,
       volume = {387},
        pages = {212},
          doi = {10.1086/171072},
       adsurl = {https://ui.adsabs.harvard.edu/abs/1992ApJ...387..212L}
}

@article{RobergClark_2016,
	author = {Roberg-Clark, G. T. and Drake, J. F. and Reynolds, C. S. and Swisdak, M.},
	doi = {10.3847/2041-8205/830/1/L9},
	journal = {Astrophys. J. Lett.},
	month = {oct},
	number = {1},
	pages = {L9},
	publisher = {The American Astronomical Society},
	title = {SUPPRESSION OF ELECTRON THERMAL CONDUCTION IN THE HIGH beta INTRACLUSTER MEDIUM OF GALAXY CLUSTERS},
	url = {https://dx.doi.org/10.3847/2041-8205/830/1/L9},
	volume = {830},
	year = {2016}}

@ARTICLE{Drake_2021,
       author = {{Drake}, J.~F. and {Pfrommer}, C. and {Reynolds}, C.~S. and {Ruszkowski}, M. and {Swisdak}, M. and {Einarsson}, A. and {Thomas}, T. and {Hassam}, A.~B. and {Roberg-Clark}, G.~T.},
        title = "{Whistler-regulated Magnetohydrodynamics: Transport Equations for Electron Thermal Conduction in the High-beta Intracluster Medium of Galaxy Clusters}",
      journal = {Astrophys. J.},
         year = 2021,
        month = dec,
       volume = {923},
       number = {2},
          eid = {245},
        pages = {245},
          doi = {10.3847/1538-4357/ac1ff1},
archivePrefix = {arXiv},
       eprint = {2007.07931},
 primaryClass = {astro-ph.GA},
       adsurl = {https://ui.adsabs.harvard.edu/abs/2021ApJ...923..245D}
}

@article{Froula_2007,
  title = {Quenching of the Nonlocal Electron Heat Transport by Large External Magnetic Fields in a Laser-Produced Plasma Measured with Imaging Thomson Scattering},
  author = {Froula, D. H., at al.},
  journal = {Phys. Rev. Lett.},
  volume = {98},
  issue = {13},
  pages = {135001},
  numpages = {4},
  year = {2007},
  month = {Mar},
  publisher = {American Physical Society},
  doi = {10.1103/PhysRevLett.98.135001},
  url = {https://link.aps.org/doi/10.1103/PhysRevLett.98.135001}
}

@article{bott_2022,
    author = {Bott, A. F. A., et al.},
    title = {Insensitivity of a turbulent laser-plasma dynamo to initial conditions},
    journal = {Matter and Radiation at Extremes},
    volume = {7},
    number = {4},
    pages = {046901},
    year = {2022},
    month = {06},
    issn = {2468-2047},
    doi = {10.1063/5.0084345},
    url = {https://doi.org/10.1063/5.0084345},
}

@article{rochau2006energy,
  title={Energy dependent sensitivity of microchannel plate detectors},
  author={Rochau, G. A., et al.},
  journal={Rev. Sci instrum.},
  volume={77},
  number={10},
  year={2006},
  publisher={AIP Publishing}
}

@article{macfarlane2006,
    series = {Radiative {Properties} of {Hot} {Dense} {Matter}},
    title = {{HELIOS}-{CR} – {A} 1-{D} radiation-magnetohydrodynamics code with inline atomic kinetics modeling},
    volume = {99},
    issn = {0022-4073},
    url = {https://www.sciencedirect.com/science/article/pii/S0022407305001627},
    doi = {10.1016/j.jqsrt.2005.05.031},
    number = {1},
    urldate = {2023-12-03},
    journal = {jqsrt},
    author = {MacFarlane, J. J. and Golovkin, I. E. and Woodruff, P. R.},
    year = {2006},
    pages = {381--397}
}

@article{Tzeferacos2017,
title = {Numerical modeling of laser-driven experiments aiming to demonstrate magnetic field amplification via turbulent dynamo},
volume = {24},
issn = {1070-664X},
url = {https://doi.org/10.1063/1.4978628},
doi = {10.1063/1.4978628},
number = {4},
journal = {Phys. Plasmas},
author = {Tzeferacos, P., et al.},
month = mar,
year = {2017},
pages = {041404},
}

\end{document}